\documentclass[useAMS,usegraphicx,usenatbib]{mn2e}
\usepackage{amsmath} 
\usepackage{subeqn, color,amssymb}

\def\be{\begin{equation}}
\def\ee{\end{equation}}
\def\bea{\begin{eqnarray}}
\def\eea{\end{eqnarray}}

\def\phidot{\dot{\phi}}

\def\pthetadot{\dot{p}_{\theta}}
\def\pthetadoti{\dot{p}_{\theta_i}}

\def\pphidoti{\dot{p}_{\phi_i}}

\def\msun{{\rm M{_\odot}}}

\def\dddot#1{\ddot#1\kern-1.4pt\dot{\phantom{#1}}\kern-3pt}

\def\dbyd#1#2{\mathchoice
             {d#1\over d#2}
             {d#1/d#2}
             {d#1\over d#2}
             {d#1\over d#2} }

\def\spose#1{\hbox to 0pt{#1\hss}}
\def\lta{\mathrel{\spose{\lower 3pt\hbox{$\mathchar"218$}}
     \raise 2.0pt\hbox{$\mathchar"13C$}}}
\def\gta{\mathrel{\spose{\lower 3pt\hbox{$\mathchar"218$}}
     \raise 2.0pt\hbox{$\mathchar"13E$}}}

\title[Self-gravitating warped discs around supermassive black holes]
{Self-gravitating warped discs around supermassive black holes}
\author[A. Ulubay-Siddiki, O. Gerhard \& M. Arnaboldi]
{A. Ulubay-Siddiki$^{1}$\thanks{E-mail: siddiki@mpe.mpg.de, gerhard@mpe.mpg.de,
 marnabol @eso.org}, 
O. Gerhard$^{1}$, and M. Arnaboldi$^{2}$
\\$^{1}$Max-Planck-Institut f\"ur Extraterrestrische Physik, 
Giessenbachstra{\ss}e, D-85748 Garching, Germany\\
$^{2}$ESO, Karl-Schwarzschild-Str. 2, D-85748 Garching, Germany}

\begin{document}



\maketitle

\label{firstpage}

\begin{abstract}
  We consider warped equilibrium configurations for stellar and
  gaseous disks in the Keplerian force-field of a supermassive black
  hole, assuming that the self-gravity of the disk provides the only
  acting torques. Modeling the disk as a collection of concentric
  circular rings, and computing the torques in the non-linear regime,
  we show that stable, strongly warped precessing equilibria are
  possible.  These solutions exist for a wide range of disk-to-black
  hole mass ratios $M_d/M_{bh}$, can span large warp angles of
  up to $\pm\sim 120\deg$, have inner and outer boundaries, and extend
  over a radial range of a factor of typically two to four.  These
  equilibrium configurations obey a scaling relation such that in
    good approximation $\phidot/\Omega\propto M_d/M_{bh}$ where
  $\phidot$ is the (retrograde) precession frequency and $\Omega$ is a
  characteristic orbital frequency in the disk. Stability was
  determined using linear perturbation theory and, in a few cases,
  confirmed by numerical integration of the equations of motion. 
    Most of the precessing equilibria are found to be stable, but some
    are unstable. The main result of this study is that highly warped
    disks near black holes can persist for long times without any
    persistent forcing other than by their self-gravity. The possible
    relevance of this to galactic nuclei is briefly discussed.
\end{abstract}


\section{Introduction}
\label{sec:intro}

The increasing power and spatial resolution of modern observations has
provided evidence that warps are not unique to galactic disks, but
appear also on much smaller scales. These include nuclear and
accretion disks surrounding supermassive black holes in galactic
nuclei (nuclear disks hereafter).  The pioneering example is the maser
disk of NGC4258.  The high velocity masers mapped by \citet{miyoshi95}
are best explained by the existence of a mildly warped disk extending
from 0.13 to 0.26 pc \citep{herrnstein}. The nearby Seyfert galaxies
NGC1068 and Circinus also harbor warped disks in their centers
\citep{greenhill03, gallimore04}, again traced by the maser emission.
{ Of the $\sim 100$ massive young stars in the center of our Galaxy
about half form a well-defined, warped disk, and some of the others
are on a counterrotating structure which may be a dissolving disk
\citep{genzel03,paumard06,lu+09,bartko08}.}

Nuclear disks can develop a warped shape through several
mechanisms. Very close to the center, the dragging of inertial frames
by a rotating black hole \citep{lense1918} causes precession of a
planar disk, if it is inclined to the plane perpendicular to the black
hole's spin.  Internal viscous torques try to align the disk angular
momentum with the black hole angular momentum. Beyond a transition
radius, the disk does not feel the effect of the black hole and
remains at its initial inclination, while inside this radius the
alignment proceeds. Hence the disk becomes warped [the
\citet{bardeen75} effect]. \citet{natarajan99} showed that for black
holes with masses of order $\rm 10^8 \ M_{\odot}$ and accretion rates
close to the Eddington limit the alignment time scale is short $(t \le
10^6 \rm{yr})$.  Application of this effect to the warped disks of
NGC4258 and NGC1068 shows that the alignment radius lies well inside
the observed positions of the maser spots, and models can be
  constructed that fit the observed warps rather well
  \citep{caproni07, martin08}.

When a warped disk is exposed to radiation from a central source, or
from its own inner portions, it is not illuminated isotropically. If
it is also optically thick, the emission received at each position and
re-radiated perpendicular to the local disk plane induces a torque on
the disk, and the warp is modified \citep{petterson77}. Perturbations
to planar disks can therefore cause radiation driven warping
\citep{pringle96}. Assuming a radiative efficiency $\epsilon \sim
10^{-2}$, and a black hole mass of $\rm 10^8 \ M_{\odot}$, an
initially flat disk is prone to warping beyond $r \ge 0.1$ pc, when
the vertical and radial viscosity coefficients are comparable
\citep{pringle97}. \cite{maloney96} studied the stable and unstable
modes of radiatively excited linear warps and found that the warp in
NGC4258 may be explained by this mechanism only if the radiative
efficieny is high.

Warps generated by gravitational interactions have been investigated 
mainly in the galactic context. \citet{hunter69} studied the
linear bending waves of a self-gravitating, isolated, thin disk.  They
showed that such a disk permits long-lasting bending modes only when
its surface density near the outer radius is truncated sufficiently
fast, but not when realistic smooth edges are considered. This
suggested interactions with nearby companion galaxies as a likely
cause of warp excitation.  Later, as the evidence for dark matter
halos around galaxies became stronger, modelers developed scenarios in
which the disk assumes the shape of a normal mode in the potential of
a flattened dark halo \citep{sparke84, Kuijken91}. However, subsequent
work showed that these modes are damped quickly when the internal
dynamical response of the halo is taken into account
\citep{nelson95,binney98}. Today, it seems most plausible that
galactic warps result from interactions and from accretion of material
with misaligned angular momentum \citep{jiang99}.

On nuclear disk scales, \cite{papaloizou98} studied in linear
  theory the evolution of a thin self-gravitating viscous disk
interacting with a massive object orbiting the central mass, with
application to NGC4258.  They concluded that the warp in the maser
disk of NGC4258 might have been excited by a binary companion with a
mass comparable to or higher than that of the maser disk. Their model
also suggests a small twist (i.e. varying line-of-nodes) due to
viscosity.  \cite{nayakshin05} considered the case of a
non-self-gravitating disk perturbed by a massive ring.  Employing the
gravitational torques in the linear regime, he evaluated the
precession induced by the ring on the disk elements. When the
self-gravity of the disk is not taken into account, the rings precess
differentially, which tends to destroy the disk structure.

Can models of warped nuclear disks be generalized to the fully
  non-linear regime? And assuming that the observed warps in galactic
  nuclei have been excited by one of the mechanisms discussed above,
  can the disk self-gravity maintain the warp even after the exciting
  torque has ceased to exist? As a first step towards answering these
  questions, the goal of the present paper is to investigate the
possibility of steadily precessing, stable, non-linearly warped
self-gravitating disks in the (Keplerian) gravitational potential of a
massive black hole. In the following sections, we use a simple
circular orbit ring model to find stable warped equilibria for systems
with 2, 3, and many rings, assuming that the self-gravity of the rings
provides the only acting torques.

\section{Steadily Precessing Warped Disks and Their Scaling Relation}
\label{sec:BH}

\subsection{Cold Disk Model and Equations of Motion}
We consider a cold disk in which stars or gas are assumed to move on
very nearly circular orbits. Following similar analysis of galactic
warps (e.g., Toomre 1983, Sparke 1984, Kuijken 1991) we model such a
disk as a collection of concentric circular rings. The orbital motion
in the disk is maintained by the central black hole, and the
self-gravity of the disk causes the rings to precess around the total
angular momentum direction.  Each ring may represent a set of stars or
gas elements uniformly spread around their circular orbit. Moreover,
when the precession frequency arising from the self-gravity of the
disk is small compared to the orbital frequency of motion, the orbital
parameters of single stars change only slowly and so one can average
over the orbital motion.  In this case, also the force exerted by a
single star or mass element on the rest of the disk can be replaced by
the force due to a ring of material spread over the orbit
\citep{goldreich}.

Any of the rings is characterized by its mass $m_i$, radius $r_i$,
inclination angle $\theta_i$ with respect to the reference plane,
and azimuthal angle $\phi_i$ where the line-of-nodes cuts this
plane. Later we will identify the reference plane as the plane
perpendicular to the total angular momentum vector. The Lagrangian 
${\mathcal L}_i$ of ring $i$ is given by
\citep{goldstein}:
\begin{eqnarray}
{\mathcal L}_i &=&  \frac{m_i r_i^{2}}{4} (\dot{\theta}_i^{2}+\dot{\phi}_i^{2} 
  \sin^{2}\theta_i) + 
  \frac{m_i r_i^{2}}{2}(\dot{\psi}+\dot{\phi}_i \cos\theta_i)^{2} \nonumber\\ 
   & & \qquad\qquad\qquad\qquad -\;V(r_i,\theta_i,\phi_i).
\label{eq:one}
\end{eqnarray}
The first two terms in equation (\ref{eq:one}) represent the kinetic
energy of the motion $T_i$, $V(r_i,\theta_i, \phi_i)$ represents the
gravitational potential energy, the Lagrangian is ${\mathcal L}_i =
T_i-V_i$, and the energy of a ring is $E_i=T_i+V_i$ .  \ $\psi$ is the
position of a point on the ring, measured from the ascending node;
($\theta,\phi,\psi$) are Euler angles.  The angular momentum of the
motion along the ring
\begin{equation}
p_{\psi_i} = m_i r_i^{2} \Omega(r_i) = m_i r_i^{2} 
      (\dot{\psi} + \dot {\phi_i} \cos\theta_i)  
\label{eq:two}
\end{equation}
is conserved since ${\mathcal L}_i$ does not depend on the coordinate
$\psi$. The other momenta are the $p_{\phi_i}$, the angular momentum
around the $z-$direction, and $p_{\theta_i}$, the angular momentum
around the line of nodes. The equations of motion are:

\begin{equation}
p_{\theta_i}=\frac{m_i r_i^{2}}{2}\dot{\theta_i},
\label{eq:three}
\end{equation}
\begin{equation}
p_{\phi_i}= \frac{m_i r_i^{2}}{2}\dot{\phi_i} \sin^{2}\theta_i +p_{\psi_i} \cos\theta_i
\label{eq:four}
\end{equation}
\begin{subequations}
 \begin{equation}
     \pthetadoti=\frac{m_i r_i^{2}}{2} \dot{\phi_i}^{2} \sin\theta_i \cos\theta_i -
     \dot{\phi_i} p_{\psi_i} \sin\theta_i -\frac{\partial V(r_i,\theta_i,\phi_i)}
       {\partial \theta_i}\\
    \label{eq:fivea}
 \end{equation}
 \begin{equation}
     \ \ =-\frac{\partial V(r_i,\theta_i,\phi_i)}{\partial \theta} -2\  
     \frac{(p_{\phi_i}-p_{\psi_i} \cos\theta_i)(p_{\psi_i}-p_{\phi_i}\cos\theta_i)}
        {m_i r_i^2 \sin^3\theta_i}
    \label{eq:fiveb}
 \end{equation}
\end{subequations}
\begin{equation}
\pphidoti=-\frac{\partial V(r_i, \theta_i, \phi_i)}{\partial \phi_i}
\label{eq:six}
\end{equation}
and the Hamiltonian is:
\begin{equation}
  \mathcal{H}_i= \frac{p_{\theta_i}^{2}}{m_i r_i^2} + 
          \frac{1}{2}\frac{p_{\psi_i}^2}{m_i r_i^2} +\frac{(p_{\phi_i}-p_{\psi_i} 
          \cos\theta_i)^{2}}{m_i r_i^{2} \sin^2\theta_i} + V(r_i,\theta_i, \phi_i).
\label{eq:seven}
\end{equation}

\subsection{Components of $V(r, \theta, \phi)$, and Evaluation of the Torques}
\label{sec:compoto}
The gravitational potential energy, $V(r, \theta, \phi)$, has two
components. One arises due to the central black hole, and is simply
\begin{equation}
V_{bh}=-\frac{G m_i M_{bh}}{r_i}.
\label{eq:eight}
\end{equation}
at the position of the ring. The other component is the potential term
$V_{m}$ arising from the interaction of the ring under consideration
with all other rings.  We follow the description of \citet{arnaboldi},
using the derivation of \citet{binney} (Section $2.6.2$), to evaluate
the torque arising from the ring interactions.

The gravitational potential due to a circular ring of mass $m_i$ and
radius $r_i$ in the $(\tilde{x},\tilde{y})$ plane is
\begin{equation}
\Phi(\tilde{x},\tilde{y},\tilde{z})=-\frac{2 G m_i}{\pi} \frac{K(k) 
\sqrt{(1-k^{2}/2)}} {\sqrt{(r^{2}+r_i^{2})}},
\label{eq:nine}
\end{equation}
where
\begin{equation}
k^{2}=\frac{4 R r_i}{(r^2+r_i^{2}+2 R r_i)}.
\label{eq:ten}
\end{equation}
Here $K(k)$ is the complete elliptic integral of the first kind, and
$R$ is the cylindrical radius $R^{2}={\tilde{x}}^{2}+{\tilde{y}}^{2}$,
so that $R^{2}=r^{2}-{\tilde{z}}^{2}$. A second ring of radius $r_j$
at an angle $\alpha_{ij}$ to the first ring follows a curve
$\tilde{z}= r_j \sin \alpha_{ij} \sin \psi$, where $\psi$ runs between $0$
and $2 \pi$. The mutual potential energy is
\begin{equation}
V_{ij}(\alpha_{ij})=-\frac{G m_i m_j}{\pi^{2}(r_i^{2}+r_j^{2})^{1/2}} \int_{0}^{2 \pi}
K(k) \sqrt{1-k^{2}/2} \hspace{0.7mm}d\psi
\label{eq:eleven}
\end{equation}
where $m_j$ is the mass of the second ring, and $k$ depends on $r_i/r_j$,
$\sin\alpha_{ij}$ and $\psi$. 
The angle $\alpha_{ij}$ between the two rings is given by
\begin{equation}
\cos\alpha_{ij} = \cos \theta_{i} \cos \theta_{j} + \sin \theta_{i} 
\sin \theta_{j} \cos (\phi_{i}-\phi_{j}),
\label{eq:twelve}
\end{equation}
which reduces to
$\cos\alpha_{ij} = \cos(\theta_{i}-\theta_{j})$
when the line-of-nodes are aligned ($\phi_i=\phi_j$). 
The torque between the two rings $(i,j)$ is
\begin{subequations}
 \begin{equation}
  \frac{\partial V_{ij}}{\partial \alpha_{ij}} 
   =  \frac{G m_{i} m_{j} r_{i} r_{j}\sin 2\alpha_{ij}}
   {   (r_{i}^{2}+r_{j}^{2})^{3/2}} \; I_{ij}(\alpha_{ij},r_i/r_j),
  \label{eq:thirteena}
 \end{equation}
 \begin{eqnarray}
  \;\; I_{ij} &\equiv& {4\over\pi^{2}}  \int_{0}^{\pi / 2}
  \left[ \frac{E(k)(1-k^{2}/2)}{(1-k^{2})} - K(k)\right] \nonumber \\
  & & \;\times \frac{(1-k^{2}/2)^{3/2}}{k^{2}} 
  \frac{\sin^2\psi\, d\psi}{\sqrt {1-\sin^{2} \alpha_{ij} \sin^2\psi}}.
  \label{eq:thirteenb}
 \end{eqnarray}
\end{subequations}
We use the numerical program of \citet{arnaboldi} for evaluating the
integrals in this expression.  The torques with respect to the angles
$({\theta_{i},\phi_{i}})$ follow from multiplying equation
(\ref{eq:thirteena}) by $\partial \alpha_{ij} / \partial \theta_{i}$
or $\partial \alpha_{ij} / \partial \phi_{i}$.  In the following, we
will write $V_{m,i}\equiv \sum_{j\ne i}{V_{ij}}$ for the potential
energy of ring $i$ due to the other rings, so that its {\sl total}
potential energy becomes
$V(r_i,\theta_i,\phi_i)=V_{bh}(r_i)+{V_{m,i}}$. For further reference
we also define $M_{ij}\equiv -\partial V_{ij}/\partial \theta_i$, and
$M_{G,i}\equiv -\partial V_{m,i}/\partial \theta_i$ for the total
gravitational torque on ring $i$ around its line-of-nodes.

\begin{figure}
\begin{centering}
\includegraphics[width=6.5cm, angle=-90]{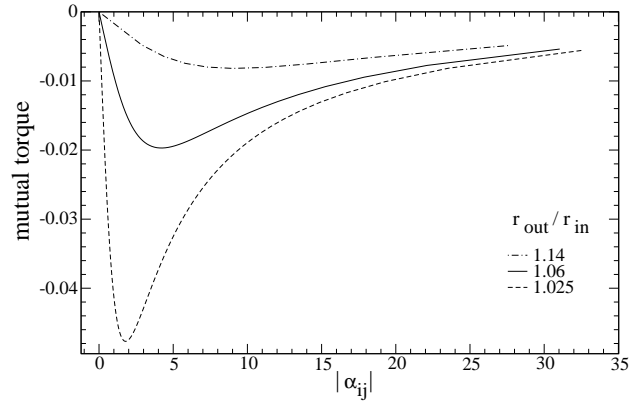}
\caption{The mutual torque between two neighboring concentric rings,
whose radii $r_{\rm out}$ and $r_{\rm in}$ are in the ratio $\nu$ as
specified on the plot. The linear regime is limited to the left
of the peak amplitude in these curves.}
\label{fig1:mtorque} 
\end{centering}
\end{figure}

Figure~\ref{fig1:mtorque} shows the torque between two rings with
radii in the ratio $\nu\equiv r_{\rm out}/r_{\rm in}$ as a function of
their mutual inclination $\alpha$. The maximum of the torque occurs at very
small angles, as noted previously by Kuijken (1991) who gives the
approximation $\alpha_{\rm max}\simeq 1.2 \vert\nu-1\vert$.  Only for
$\alpha<\alpha_{\rm max}$ can the mutual torque be approximated as a
linear function of $\alpha$. Thus solutions $\theta(r)$ for the warp
shape in linear theory can be scaled by a constant multiplicative
factor only so long as the local gradient $\dbyd{\theta}{r}<1.2/r$.
Otherwise the local self-gravity torques of the disk are no longer
able to maintain the linear theory warp shape, the non-linear
equations must be used, and the shape of the warp must change.

\subsection{Steadily Precessing Equilibria}
\label{prequi}

A configuration of inclined rings precessing as a rigid body with
constant $\dot{\phi}$ will in the following be denoted as a steadily
precessing equilibrium, or equilibrium for short.  In earlier works by
\citet{hunter69,sparke84}, and \cite{sparke88} it was found that the
eigenfrequencies of linear m=1 warp modes are purely
real. \citet{papaloizou98} showed that this is a consequence of the
self-adjointness of the operator in the tilt equation when there are
no viscous or non-conservative forces.  Then the eigenvectors are also
real and thus the warp has a straight line-of-nodes and no
spirality\footnote{ The line-of-nodes for a set of orbits is here
  defined as the union of points where all inclined orbits
  ($\theta_i\ne 0$) with respect to the reference plane intersect this
  plane.}.  In the light of these linear theory results our effort
will also be to find equilibria where all the rings have the same
azimuth $\phi$. The condition that all rings maintain constant
inclination, $\dot\theta=0$, implies also $\pthetadot=0$, and for a
given precession rate $\dot{\phi}$, the simultaneous solution of this
equation for each of the rings determines the inclination angles,
i.e. the equilibrium shape corresponding to this value of
$\dot{\phi}$.  We note that for $\dot{\phi}=0$
equations~(\ref{eq:fivea}) admit a trivial tilt solution
$\theta_i={\rm const.}$ but will assume $\dot{\phi}\ne0$ in what
follows.

From equations (\ref{eq:fivea}) and (\ref{eq:two}) we can solve for
the precession rate of ring $i$:
\begin{eqnarray}
&&\dot{\phi_i} = \frac{\Omega_{i}}{\cos \theta_{i}} \pm \nonumber \\
&& \sqrt{   \frac{\Omega_{i}^2}{\cos^2\theta_{i}} +
      \frac{2}{ m_{i} r_{i}^2 \cos \theta_{i} \sin \theta_{i} }
      \displaystyle\sum_j{\frac{\partial V_{ij}}{\partial \theta_{i}}}}
\label{eq:fourteen}
\end{eqnarray}
when $\theta_i\ne 0$.  Here $\Omega_i=\sqrt{G M_{bh}/r_i^{3}}$ is the
angular velocity of particles on the ring around the black hole, and
the term $\sum_j \partial V_{ij}/ \partial \theta_{i}$ is the torque
on ring $i$ caused by all other rings $j$. The precession rate can
therefore be fast or slow, corresponding to the plus and minus signs
in this expression. When the interaction potential $V_{ij}$ increases
away from the plane $\theta=0$, the second term in the square root is
positive, so that the slow precession is retrograde
($\dot{\phi}<0$). In the remainder of this paper we focus on such slow
retrograde precession.

The components of angular momentum along the original $(x,y,z)$ axes
for a single ring read: 
\begin{equation}
l_{x_i} = p_{\theta_i} \cos\phi_i + 
        \frac{\sin\phi_i(p_{\psi_i}-p_{\phi_i}\cos\theta_i)}{\sin\theta_i},
\label{eq:fifteen}
\end{equation}
\begin{equation}
l_{y_i} = p_{\theta_i} \sin\phi_i - 
        \frac{\cos\phi_i(p_{\psi_i}-p_{\phi_i}\cos\theta_i)}{\sin\theta_i},
\label{eq:sixteen}
\end{equation}
\begin{equation}
l_{z_i} = p_{\phi_i}.
\label{eq:seventeen}
\end{equation}
Let us assume that we have found a precessing equilibrium from solving
equations~(\ref{eq:three}-\ref{eq:six}), with $p_{\theta_i}=0$,
$\dot{p}_{\theta_i}=0$, and $p_{\phi_i}={\rm const.}$,
$\dot{\phi}={\rm const.}$, ${\phi_i}=\phi$.  Inserting
equations~(\ref{eq:fivea}) and (\ref{eq:four}) into the expression for
$l_{x_i}$, simplifying and summing over all rings gives the total
angular momentum
\begin{equation}
l_x=\displaystyle\sum_i{l_{x_i}}=-\frac{\sin\phi}{\dot{\phi}}
    \displaystyle\sum_{i\ne j}\frac{\partial}{\partial \theta_i} V_{ij} =0
\label{eq:eighteen}
\end{equation}
which sums to zero because for each pair of rings with interaction
potential $V_{ij}$ the torques are equal and opposite. Similarly, the
total $l_y=0$. Thus the total angular momentum of such a precessing
equilibrium configuration is parallel to the $z$-axis. By construction, 
the angular momentum of the precession alone is also along the $z$-axis,
{ i.e., the disk precesses around the total angular momentum vector axis.}

For a uniformly precessing configuration, additional insight may be
obtained by moving to a coordinate system which rotates around the
angular momentum axis with the disk's precession frequency
$\dot{\phi}$ (Kuijken 1991). In this reference frame the shape of the
precessing disk is stationary, but the particles in the different
rings still spin about their rings' symmetry axes. If the particles in
ring $i$ rotate with velocity $\Omega(r_i) r_i$ in the positive sense,
they experience a Coriolis force in the rotating system which,
integrated over the ring, results in a Coriolis torque on ring $i$
along the $p_{\theta}$-axis (line-of-nodes), given by
\begin{equation}
M_{C,i} = -m_i r_i^2 \Omega(r_i) \dot{\phi}\sin\theta_i.
\label{eq:nineteen}
\end{equation}
For $0<\theta<{\pi/2}$ and negative $\dot{\phi}$ this torque is along
the positive $p_{\theta}$-axis, i.e., is trying to retard the ring
relative to the rotating frame.  Because the retrograde precession
speeds are small, we can neglect the centrifugal force terms. In this
case, a stationary precessing configuration is obtained when the
forward gravity torques and the retarding Coriolis torque balance in
the rotating frame.

\subsection{2-Ring and 3-Ring Cases}
\label{sec:threerings}
The argument just described suggests that there should exist steadily
precessing 2-ring configurations in which one ring is tilted above the
plane $\theta=0$ and a second ring is tilted below this plane. Both
rings are pulled towards $\theta=0$ by the gravitational force from
the other ring. The resulting gravity torques cause the angular
momentum vectors of the two rings to precess in the same sense, and
are balanced by the Coriolis torques in the precessing frame. To find
such configurations we need to solve $\pthetadot=0$ using
eq.~(\ref{eq:fivea}) for both rings simultaneously. Assuming
$\phidot\ll\Omega$, we can neglect terms of order $\phidot^2$; then
using eq.~(\ref{eq:two}) the equation for the inner ring at $r_1$
becomes
\begin{equation}
\sin\theta_1\simeq -\frac{\partial V_{12}}{\partial\theta_1}
    / m_1 r_1^2 \Omega(r_1) \phidot
\label{eq:twenty}
\end{equation}
and the ratio of the two equations is
\begin{equation}
\sin\theta_1/\sin\theta_2\simeq -m_2 r_2^2 \Omega(r_2) / 
                                 m_1 r_1^2 \Omega(r_1),
\label{eq:twentyone}
\end{equation}
where $m_1$, $m_2$ are the two ring masses, $r_1$, $r_2$ their
radii, $\theta_1$, $\theta_2$ their inclinations, and $V_{12}$ the
interaction potential. Given the ring masses and radii and $\theta_1$,
say, we can determine from these equations $\theta_2$, the interaction
potential, and thus finally the precession rate $\phidot$ required
for steady precession.

Using the expression in (\ref{eq:thirteena}) for the torque between the
rings, equation (\ref{eq:twenty}) can be cast into a more useful
form:
\begin{equation}
\frac{\dot{\phi}}{\Omega_{1}}=-\frac{\sin 2\alpha\;I_{12}(\alpha,\nu)}{\sin \theta_1} 
           \frac{\nu\mu}{(1+\nu^2)^{3/2}} \;
           \frac{m_{1}}{M_{bh}},
\label{eq:twentytwo}
\end{equation}
where the angle $\alpha=\theta_1-\theta_2$, $\nu\equiv r_2/r_1$, $\mu= m_2/m_1$,
and $I_{12}$ denotes the integral expression of equation (\ref{eq:thirteenb}).

\begin{figure}
\begin{centering}
\includegraphics[width=8cm, angle=0]{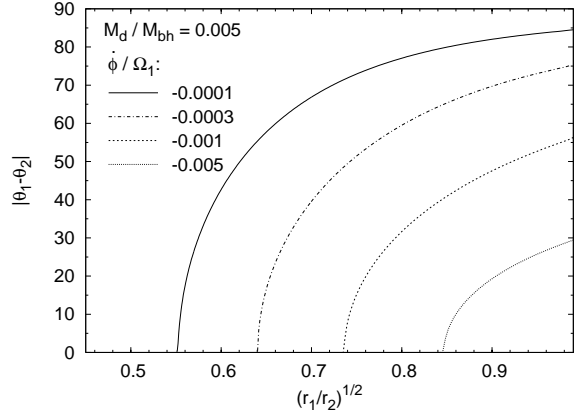}
\caption{Relative inclination of two rings precessing together around 
a central mass $M_{bh}$, as a function of the square root of the ratio of their
radii. The combined mass of the two rings is $0.5\%$ of the central mass,
approximately as in the Galactic center. The different curves are for 
various values of the precession frequency of the two-ring system, in
units of the Keplerian frequency at $r_1$.}
\label{fig2:2rings}
\end{centering}
\end{figure}

Figure \ref{fig2:2rings} shows the difference $\theta_1-\theta_2$
between the inclination angles of the two rings versus the square root
of the ratio of their radii, $r_1/r_2$, for different precession
frequencies, expressed in units of the Keplerian frequency at
$r_1$. The combined mass of the two rings is chosen to be $0.5\%$ of
the central mass, approximately as inferred for the system of two
stellar rings in the Galactic center
\citep{genzel03}. $\theta_1-\theta_2$ increases with decreasing
precession speed when the mass ratio is
fixed. Eq.~(\ref{eq:twentytwo}) shows that the same precessing
equilibrium configuration can be obtained by changing
$\dot{\phi}\propto M_d=m_1+m_2$ and leaving all other parameters
unchanged. More massive rings must precess faster for the same
inclinations. Thus the sequence of curves in Fig.~\ref{fig2:2rings}
can also be interpreted as a sequence of fixed precession frequency
but with mass ratio $M_d/M_{bh}$ increasing from bottom right to upper
left.
 
\begin{figure}
\begin{centering}
\includegraphics[width=6cm, angle=-90]{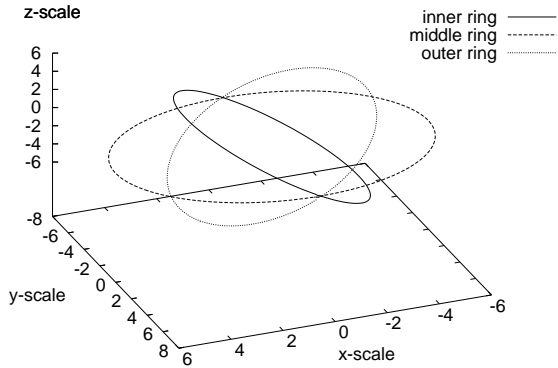} 
\caption{3D view of a 3-ring system. The middle ring lies close to the
equator, while the others are distributed almost symmetrically around
it.}
\label{fig:threeD3rings} 
\end{centering}
\end{figure}

Next consider three rings. In this case, each of the rings precesses
in the potential of the other two rings, and the reference frame is
defined by the common plane of precession perpendicular to
the total angular momentum vector. Again $\pthetadot$
[eq.~(\ref{eq:fivea})] should be zero at equilibrium for each of the
rings. We can sum these three equations:
\bea
\displaystyle\sum_{i}^{3} \dot{p}_{\theta_i} &=&
     \displaystyle\sum_{i=1}^{3} \left( 
        \frac{m_{i} r_{i}^2}{2}  \dot{\phi}^2
           \sin{\theta_{i}} \cos{\theta_{i}}
        - \dot{\phi} p_{\psi_i} \sin{\theta_{i}} \right) \nonumber \\
      &{}& \qquad\qquad\qquad
        - \displaystyle\sum_{i=1}^{3} \displaystyle\sum_{j\ne i} 
        \frac{\partial V_{ij}}{\partial \theta_i}=0.
\label{eq:twentythree}
\eea
The $V_{ij}$ terms cancel since 
$\partial V_{ij}/\partial\theta_i=-\partial V_{ij}/\partial\theta_j$. 
The remaining terms can be rewritten as
\begin{equation}
    \dot{\phi} \left( \, \displaystyle\sum_{i=1}^{3} 
       m_{i} r_{i}^2\sin\theta_{i}
       \left[ -\Omega(r_{i})+\frac{1}{2}\dot{\phi}\cos\theta_{i}\right] 
               \right) = 0,
\label{eq:twentyfour}
\end{equation}
making use of eq.~(\ref{eq:two}). This shows that, apart from the
no-precession solution, a steadily precessing equilibrium is possible
only when at least one of the rings lies on the opposite side of the
equator with respect to the others, i.e., has $\theta_i<0$. Likewise
the two rings of a precessing two-ring system must lie on opposite
sides of the equator. Figure \ref{fig:threeD3rings} shows as an example the 
3D view of a 3-ring system with mass $M_d = 0.05 M_{bh}$ and 
$\dot{\phi}=-0.0021\Omega(r_2)$.

\subsection{Approximation of A Disk With n-Rings}
\label{sec:nringapprox}

\begin{figure}
\includegraphics[angle=0,width=8cm]{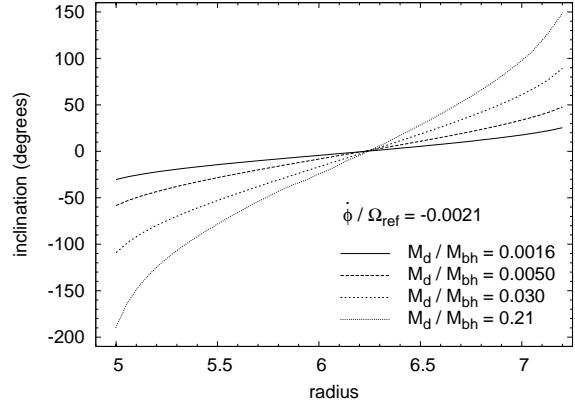}
\centering
\caption{Inclination of a disk of constant surface density at
  different radii. The model consists of 35 rings precessing at a rate
  of $\phidot=-10^{-3}$ units, so that $\phidot/\Omega(r_{\rm
    ref})=-0.0021$ on its middle (reference) ring. Each curve
  corresponds to a different $M_d/M_{bh}$ mass fraction.  The warp
  becomes more pronounced when the disk mass is increased. The
  smallest and highest masses correspond to the limits for stability
  (see Section \ref{sec:anstab}).}
\label{fig4:evalcsd}
\end{figure}

We now consider a disk represented as a collection of $n$ concentric
rings. To find a precessing equilibrium, we solve $\dot{p}_\theta=0$
(eqs.~\ref{eq:fivea}) for all rings simultaneously, summing over the
torques from all other rings (eqs.~\ref{eq:thirteena}).  These
  are $n$ equations for n+1 unknowns, the n inclinations $\theta_i$
  and $\dot{\phi}$, which we solve keeping $\dot{\phi}$ fixed
  \citep{arnaboldi}\footnote{Note that this does not work in linear
    theory because the linear solution can be scaled arbitrarily,
    i.e., one of the $\theta_i$ can be eliminated.}.  Figure
\ref{fig4:evalcsd} shows a sequence of equilibria obtained for a
constant surface density disk consisting of $35$ rings. On each curve,
the extent of the disk $(i.e. \ \Delta r = r_{\rm out}-r_{\rm in})$ is
fixed at $2.2$ units, and the precession rate is
$\phidot/\Omega(r_{\rm ref})=-0.0021$ where $\Omega(r_{\rm ref})$ is
the circular frequency on the middle (reference) ring. The disk mass
fraction $M_d/M_{bh}$ varies from $0.16\%$ to $21\%$. As the mass of
the disk increases, the degree of warping increases dramatically {
  so that the Coriolis torques can keep the balance of the gravity
  torques.} The basic shape of the disk is similar to that of the
system of three rings in Fig.~\ref{fig:threeD3rings}. The middle rings
lie closest to the equator, while the inner and outer rings are almost
symmetrically distributed around it.
%

\begin{figure}
\begin{centering}
\includegraphics[width=6cm, angle=-90]{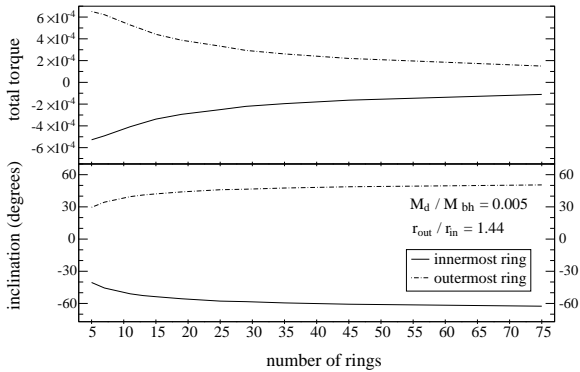} 
\caption{Convergence of the total torques (upper panel), and
inclinations (lower panel), for the innermost (solid lines) and
outermost rings (dot-dashed lines), when the number of rings to
represent the disk is increased but the extent and the mass of the
disk are kept fixed.}
\label{fig5:nringvsth} 
\end{centering}
\end{figure}

Obviously, the larger the number of rings the better the approximation
to a continuous disk. Figure~\ref{fig5:nringvsth} shows the
convergence of the total torques (upper panel), and of the inclination
angles obtained in steadily precessing equilibrium (lower panel), for
the innermost and outermost rings, when the number of rings to
represent the disk is increased but the extent and the mass of the
disk are kept fixed. One sees that quite a number of rings
are needed before the torques converge. The inclination of the 
outer and inner rings have approximately converged when $n\gta 30$.

\begin{figure}
\includegraphics[width=6cm, height=9cm, angle=-90]{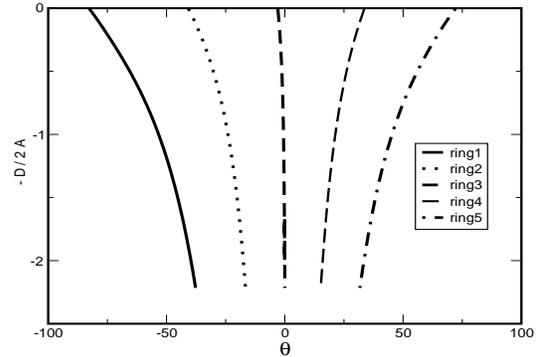} 
\includegraphics[width=6cm, height=9cm, angle=-90]{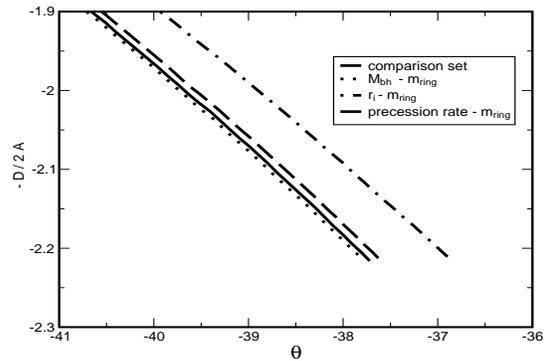} 
\centering 
\caption{Top: Values $-D_{i} / 2 A_{i}$ (eq.~\ref{eq:twentyeight}) for
  a system of $5$ rings with the parameters described in the text.
  The absolute value of the precession speed increases along the
  curves from top to bottom.  Bottom: Parameter scaling of this system
  according to eq.~(\ref{eq:twentyeight}).  This figure shows a zoom
  into the region where the orientation of the $1^{st}$ ring predicted
  by the scaling deviates from the orientation in the original 5 ring
  model (shown as the solid line). Dots show when the black hole mass
  is increased by a factor of 4 and the ring mass is increased by a
  factor of 2, the dotted-dashed line shows when the radius is
  increased by a factor of 4 and ring mass by a factor of 8, and the
  long dashed line shows when the scaling is done by increasing the
  precession rate by a factor of 2 while increasing again the ring
  mass by a factor of 2.}
\label{fig:orig1BW}
\end{figure}

\subsection{Scaling the Solutions}
\label{sec:scaling}
Now we go back to the equilibria themselves, in particular to the
question of their scaling properties. When
the torque on ring $i$ from all other rings is decomposed as
\begin{eqnarray}
    M_{G,i} = - \displaystyle\sum_{j} \frac{\partial V_{ij}}{\partial \theta_{i}}
            = - \displaystyle\sum_{j} \frac{\partial V_{ij}}{\partial \alpha_{ij}}
                               \frac{\partial \alpha_{ij}}{\partial \theta_{i}},
\label{eq:twentyfive}
\end{eqnarray}
the mass- and radius-dependent part is the first derivative on the
r.h.s. The second factor in each term of this sum depends from
equation~(\ref{eq:twelve}) only on the two sets of angles
$\theta_i,\phi_i$, $\theta_j,\phi_j$. For equilibria with a common
precessing line-of-nodes, $\alpha_{ij}=(\theta_i-\theta_j)$, so the
derivative is always unity. For the potential derivative terms, we use
equation (\ref{eq:thirteena}) and express all ring masses and radii in
terms of the mass and radius of a reference ring, i.e., we write
$\mu_i\equiv m_i/m_{\rm ref}$, $\nu_i\equiv r_i/r_{\rm ref}$, and
similar for $j$.  Then (\ref{eq:twentyfive}) takes the form:
\begin{equation}
  M_{G,i} =   - \frac{Gm_{\rm ref}^2}{r_{\rm ref}} \underbrace{
    \left[\displaystyle\sum_{j} {\frac{\mu_{i} \mu_{j} \nu_{i} \nu_{j} \sin\alpha_{ij}}
         {(\nu_{i}^2+\nu_{j}^2)^{3/2}} I_{ij}(\alpha_{ij},\nu_i,\nu_j)} \right] }
\label{eq:twentysix}
\end{equation}
where we denote the expression over the brace as $D_i/2$. For each precessing
equilibrium disk configuration as in Fig.~\ref{fig4:evalcsd} $D_i$ is a constant,
and the torques on all rings scale as $\propto m_{\rm ref}^2/r_{\rm ref}$.
If we now go back to equation (\ref{eq:fourteen}), insert equation (\ref{eq:twentysix}),
and normalize the precession rate with the circular frequency at the reference
radius, $\Omega_{\rm ref}\equiv(GM_{bh}/r^3_{\rm ref})^{1/2}$, 
we find
\begin{eqnarray}
\frac{\dot{\phi_i}}{\Omega_{\rm ref}} = \frac{1}{\nu_i^{3/2}\cos\theta_i} 
  \hspace{1mm}\left( 1 \pm \sqrt{1+
     \frac{\nu_i \cos\theta_i}{\mu_i \sin\theta_i} 
         \; D_i \; \frac{m_{\rm ref}}{M_{bh}} }\; \right).
\label{eq:twentyseven}
\end{eqnarray} 
For negative $\phidot$ and after a Taylor expansion of the
term in the square root, appropriate for slow retrograde precession,
equation~(\ref{eq:twentyseven}) becomes
\begin{eqnarray}  
\frac{\dot{\phi_i}}{\Omega_{\rm ref}} 
     \simeq - \frac{D_{i}}{2\mu_i\nu_i^{1/2}\sin\theta_i} 
            \; \frac{m_{\rm ref}}{M_{bh}}
     \equiv - \frac{D_{i}}{2 A_i} \; \frac{m_{\rm ref}}{M_{bh}}.
\label{eq:twentyeight}
\end{eqnarray}
A precessing equilibrium is one for which all rings precess with the
same common frequency, $\dot{\phi_i} = \dot{\phi}$. Equation
(\ref{eq:twentyeight}) thus shows that for a fixed precessing disk
mass configuration (i.e., fixed ring masses, radii, and inclinations,
hence fixed $D_i/2A_i$), the precession rate $\dot{\phi}$ scales
proportional to the Keplerian frequency at some reference radius in
the disk and proportional to the disk-to-black hole mass ratio.  Vice
versa, equation (\ref{eq:twentyeight}) can be interpreted as a scaling
relation which says that a precessing equilibrium solution remains
unchanged in shape ($\theta_i$) under changes of the disk mass, disk
radius, black hole mass, and precession rate, provided the ratio
$(\dot{\phi}/\Omega_{\rm ref})/(M_d/M_{bh})$ is held constant.

Figure \ref{fig:orig1BW} depicts the values of $-D_{i}/2A_{i}$ for a
system of $\rm 5$ rings.  The radii of the rings are calculated such
that $r_{i}=\kappa^{i-1}\times r_{1}$, with $i=1,2,..n$,
$\kappa=1.07$, $r_{1}=5.75$, and $n=5$, so if the third ring is the
reference ring, $\nu_{i}=\kappa^{i-3}$.  The ring masses are assumed
to all have the same value, $0.5516$, so $\mu_i=1$, the black hole has
a mass of 51.16, and $M_d/M_{bh}= 0.054$. On each curve in
Fig.~\ref{fig:orig1BW}, the precession rate $\dot{\phi}$ increases
with steps of $-5\times10^{-5}$
($\Delta\phidot/\Omega(r_3)=-1.18\times10^{-4}$) starting from 
a value of $-1\times10^{-4}$
(i.e., $\phidot/\Omega(r_3)=-2.4\times10^{-4}$) at the top. We checked the
accuracy of the scaling and of our calculations by computing
$-D_{i}/2A_{i}$ values for different parameter pairs of the system
that should give the same $-D_{i}/2A_{i}$ according to
equation~(\ref{eq:twentyeight}). We overlay the results for the
first ring and precession speed $\phidot/\Omega(r_3)=-0.0021$, 
in the lower panel of Fig.~\ref{fig:orig1BW}, zooming into
the parameter region $-2.3 < D_{1}/2A_{1} <-1.9$ where the different
curves deviate from each other the most. In the worst case, due to the
scaling of the ring radii, the deviations of the $\theta_i$'s from
their values for the original 5 ring system are still less than $\rm
1^{\circ}$. Changes in radii cause the largest deviations from the
scaling relation because of the way in which they enter in the
quantity $D_i$ (equation~\ref{eq:twentyseven}).  The scaling results
for the other rings are similar.

\subsection{Stability}
\label{sec:anstab} 

{ In this section we investigate the stability of the precessing
  equilibrium solutions found above.  \cite{hunter69} proved that
  isolated thin self-gravitating disks are stable to all $m=1$ warp
  perturbations and this carries over to disks embedded in spherical
  or oblate potentials \citep[e.g.][]{sparke88}. We show here that the
  non-linearly warped precessing disks can be both stable and unstable
  to general ring-like perturbations. We describe small perturbations
  of the precessing disk solutions} by the linearized equations of
motion around equilibrium:
\begin{eqnarray}
  \Delta\dot{\theta_{i}}&=& \frac{\partial^2 T_{i}}{\partial p_{\theta_{i}}^2 } 
				\Delta p_{\theta_{i}},\nonumber \\
  \Delta\dot{\phi_{i}}  &=& 
		    \frac{\partial^2 T_{i}}{\partial p_{\phi_{i}}\partial{\theta_{i}}}
   				 \Delta \theta_{i} 
			    + \frac{\partial^2 T_{i}}{\partial p_{\phi_{i}}^2} 
				 \Delta p_{\phi_{i}},\nonumber \\
  \Delta\dot{p}_{\theta_{i}} &=& 
		    -\frac{\partial^2 T_{i}}{\partial\theta_{i} \partial p_{\phi_{i}}}
 				 \Delta p_{\phi_{i}} 
		    -\frac{\partial^2 T_{i}}{\partial\theta_{i}^2}
 				 \Delta \theta_{i}
		    -\frac{\partial^2 V_{m,i}}{\partial\theta_{i}^2} 
				 \Delta \theta_{i}\nonumber\\
		    & & -\frac{\partial^2 V_{m,i}}{\partial\theta_{i} \partial\phi_{i}} 
				 \Delta \phi_{i} 
		    -\sum_j\frac{\partial^2 V_{m,i}}{\partial\theta_{i}\partial\phi_{j}} 
				 \Delta \phi_{j}
		    -\sum_j\frac{\partial^2 V_{m,i}}{\partial \theta_{i}\partial \theta_{j}} 
				 \Delta \theta_{j}, \nonumber\\
  \Delta\dot{p}_{\phi_{i}} &=& 
		    -\frac{\partial^2 V_{m,i}}{\partial\phi_{i} \partial\theta_{i}} 
				 \Delta\theta_{i}
		    -\frac{\partial^2 V_{m,i}}{\partial\phi_{i}^2} 
				 \Delta \phi_{i} \nonumber\\
		    & & -\sum_j\frac{\partial^2 V_{m,i}}{\partial\phi_{i} \partial\theta_{j}} 
				 \Delta\theta_{j}
		    -\sum_j\frac{\partial^2 V_{m,i}}{\partial\phi_{i} \partial\phi_{j} } 
				 \Delta\phi_{j}.
\label{eq:thirty}
\end{eqnarray}
Here $T_i$ and $V_{m,i}$ are the kinetic and potential energy terms in
the Hamiltonian (\ref{eq:seven}) for ring $i$, respectively, and the
partial derivatives are evaluated at the equilibrium solution
$(\theta_i,\phi_i\!=\!{\rm const.},\dot{p}_{\theta_i}\!=\!0,
\dot{p}_{\phi_i}\!=\!0)$. These linear equations have solutions
of the form $\rm{e}^{\lambda t} \Delta\theta_0, \dots$ etc., where
$\lambda=\lambda_{\mathcal{R}}\pm \rm{i} \lambda_{\mathcal{I}m}$ with its
real and imaginary parts.  The (constant) coefficients of the $\Delta$
terms in equation (\ref{eq:thirty}) form a matrix H which carries the
information on stability. When the real parts of the eigenvalues of
the matrix H, $ \lambda_{\mathcal{R}} = 0$, the imaginary parts of the
eigenvalues, $\lambda_{\mathcal{I}m}$, constitute a rotation matrix
through which the solutions oscillate around the precessing
equilibrium with frequencies ${\lambda_{\mathcal{I}m}}$, and the
equilibrium is said to be stable. When $ \lambda_{\mathcal{R}} < 0$,
the solutions spiral towards the unperturbed equilibrium positions,
leading to asymptotic stability of the equilibrium. If, however, any
of the eigenvalues have a nonzero real part, $\lambda_{\mathcal{R}} >
0$, the system moves away from equilibrium exponentially, and is
unstable.

For determining the stability of any of our precessing $n$-ring solutions,
we compute the $4n\times 4n$ stability matrix $H$, using the equilibrium 
$(\theta_i,\phi_i\!=\!{\rm const.},\dot{p}_{\theta_i}\!=\!0,
\dot{p}_{\phi_i}\!=\!0)$. We then evaluate the eigenvalues of the
matrix H, using routine F02EBF of the Numerical Algorithms
Group (NAG). This routine is suitable for computing eigenvalues and
optionally eigenvectors of real matrices.

\begin{figure}
\centering
\includegraphics[angle=-90,width=8cm]{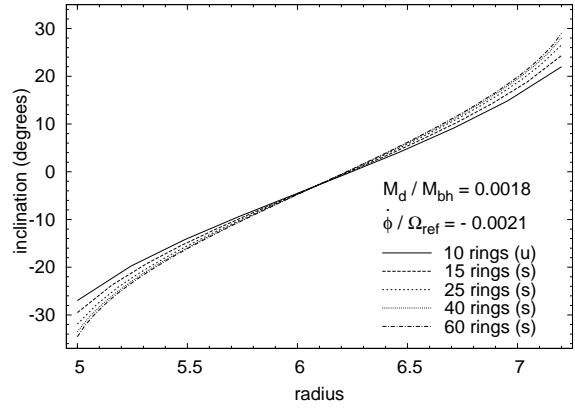}
\caption{Change of the shape of a solution near the lower stable boundary of
  Fig.~\ref{fig4:evalcsd} with the number of rings used to represent
  the disk. The disk-to-black hole mass ratio, precession speed, and
  number of rings are given on the plot. The letters ``u'' and ``s''
  in parentheses following the number of rings stand for unstable and
  stable solutions respectively. }
\label{fig:convergence}
\end{figure}

{
First, we briefly discuss one example for the convergence of the
linear stability results.  In Section~\ref{sec:nringapprox} we had
already discussed the convergence of the gravitational torques, and of
the inclination angles obtained for the precessing equilibrium
solutions, as a function of the number of rings used to represent the
disk (see Fig.~\ref{fig5:nringvsth}).  Figure~\ref{fig:convergence}
shows how the shape of a solution near the lower stable boundary of
Fig.~\ref{fig4:evalcsd} and its stability changes with the number of
rings used to represent the disk. The transition from unstable to
stable occurs at a disk mass fraction of $M_d/M_{bh}=0.0018,
0.0013,0.0011,0.0011$ for $n=15,25,45,75$ rings.  This shows that the
transition has approximately converged when $n\simeq45$.
}

\begin{figure}
\centering
\includegraphics[angle=0,width=8cm]{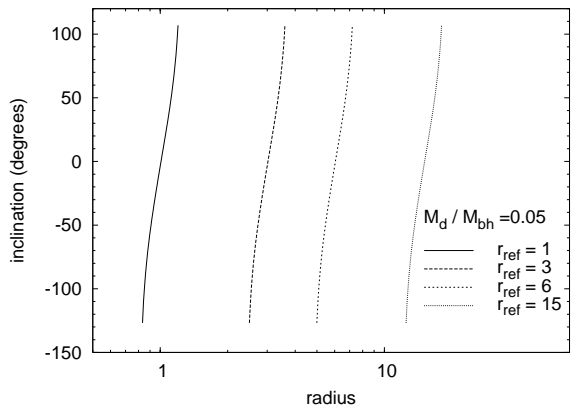}
\caption{35-ring-disks with constant $M_{d}/M_{bh}=0.05$ but different sizes.
The curves from left to right show warped disk solutions for which the
scale radius $r_{\rm ref}$ is shifted to ever higher values while
correspondingly $\dot{\phi}$ is decreased to keep
$\dot{\phi}\sqrt{r_{\rm ref}^3}$ fixed.  The innermost and outermost
ring inclinations are identical for all disks, as expected from the
scaling relation, and all rescaled disks are found to be stable.}
\label{fig:constbeta}
\end{figure}

{
Next, we consider the issue of scaling. We have already seen that the
equilibrium solutions can be scaled in radius, mass, and precession
speed according to the approximate (but accurate) scaling relation
(\ref{eq:twentyeight}): the angles $\theta_i$ for all rings in the
disk remain unchanged if the precession speed expressed in units of
the angular frequency scales linearly with the disk-to-black hole mass
fraction. Figure \ref{fig:constbeta} shows disks with
$M_{d}/M_{bh}=0.05$ consisting of equal mass rings with constant
ratios of all ring radii relative to each other.  Keeping the ratio
$\dot{\phi} \sqrt{r_{\rm ref}^3}$ constant, we have moved the
midpoints of these disks to various radii. All these configurations
are stable, and actually represent a similar warped disk (i.e.
innermost and outermost inclinations being the same) at different
distances from the black hole.
}

\begin{figure}
\centering
\includegraphics[width=8cm]{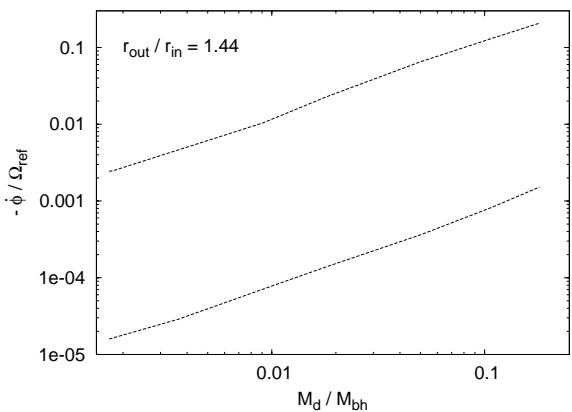}
\caption{Schematic representation of stable warped disk solutions in
  the $M_d/M_{bh}$ versus $\phidot/\Omega$ plane. The two inclined
  dashed lines { enclose the stable region: the upper (lower) dashed
  line corresponds to the maximum (minimum) stable precession
  frequency at fixed disk-to-black hole mass ratio. }  This diagram is
  for disks with the same surface density profile and inner and outer
  boundaries $r_{\rm in}$, $r_{\rm out}$.}
\label{fig:schematic}
\end{figure}

{ 
Figure \ref{fig:schematic} illustrates the stability properties for
these disks when the disk mass and precession speed are changed
simultaneously.  The boundaries of the stable solutions obtained for a
fixed disk mass configuration, i.e., a configuration with constant
ratios of all ring masses and radii relative to each other, are shown
by the two dashed lines in the figure. All stable solutions lie
between the two lines. These dashed lines are nearly straight,
indicating that the minimum and maximum $\phidot/\Omega_{\rm ref}$
solutions at different mass fractions can be essentially scaled to
eachother - if the same (scaled) equilibrium solution was the
stability boundary for all mass ratios, the lines would reflect the
scaling relation of eq.~(\ref{eq:twentyseven}) respectively 
eq.~(\ref{eq:twentyeight}) precisely.
}

{ Figure \ref{fig:schematic} shows that for each ratio of disk mass
  to black hole mass, there is a minimum and maximum stable precession
  speed $\phidot/\Omega$, and vice-versa.  (see
  Fig.~\ref{fig4:evalcsd} and Section~\ref{sec:warp} below)}. The
minimum (maximum) stable mass for given $\phidot/\Omega$ is stable for
all precession speeds lower (higher) than the original
$\phidot/\Omega$, until the other boundary curve is reached. For other
disk mass configurations, the range of stable mass ratios and the
corresponding boundary lines change, but the qualitative behaviour
remains the same.

\section{Steadily Precessing, Non-linearly Warped Keplerian Disks: Results}
\label{sec:results}

\subsection{Warp Shapes and Warp Angles of Stable Precessing Disks}
\label{sec:warp}
%
\begin{figure}
\centering
\includegraphics[angle=-90,width=8cm]{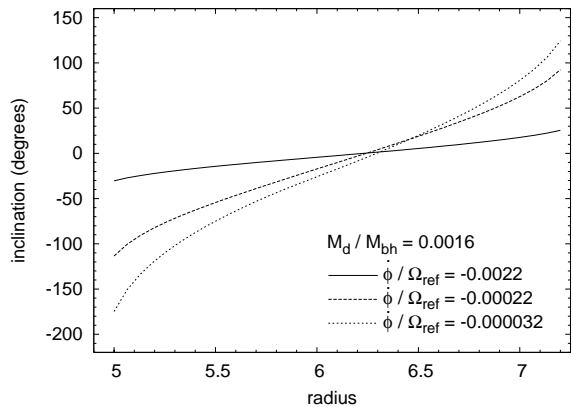}
\caption{Inclination of a disk of 35 rings at different radii for
$M_{d}/M_{bh}=0.0016$. On each curve the precession frequency
$\dot{\phi}$ has a different value, given on the figure in terms of
the orbital frequency at the position of the middle ring. The upper
and lower curves show the boundaries of stable solutions. See
Section~\ref{sec:anstab} and Fig.~\ref{fig:schematic}. }
\label{fig:varpdtm00018}
\end{figure}

\begin{figure}
\centering
\includegraphics[angle=0,width=8cm]{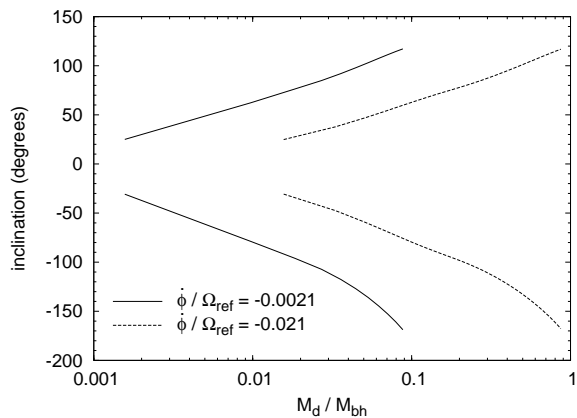}
\caption{Variation with disk-to-black hole mass ratio of the
  inclinations of the outermost $(\theta>0)$ and innermost
  $(\theta<0)$ rings of a 35-ring warped disk with $r_{\rm out}/r_{\rm
    in}=1.44$, for two different values of the precession frequency.
  The figure shows how the warping increases with increasing disk mass
  fraction and with decreasing precession speed, and also illustrates
  the scaling relation of equation~(\ref{eq:twentyeight}).  Up to
  numerical errors, the minimum and maximum values of $\theta$ are the
  same on all two curves.}
\label{fig:fixphdt}
\end{figure}

We have already seen in the discussion of three-ring and n-ring
systems in Section~\ref{sec:BH} that self-gravitating precessing disks
in a Keplerian potential can be strongly warped. { In fact, some of
  the disks shown in Fig.~\ref{fig4:evalcsd} are so strongly warped
  that they would obscure the central black hole from most
  lines-of-sight.}

In this section we discuss these results in more
detail. Figure~\ref{fig:varpdtm00018} shows the warped stable
equilibrium solutions obtained for a sequence of disks with varying
precession frequency. These disks have constant surface density
between fixed inner and outer radii, and a total disk-to-black hole
mass ratio of $M_{d}=0.0016M_{bh}$.  The solutions shown in
Fig.~\ref{fig:varpdtm00018} are all linearly stable, according to the
analysis described in Section~\ref{sec:anstab}.  Outside the range of
models bounded by the upper and lower curves one can find further
equilibria, but these are unstable.

By construction, these disks have a fixed line-of-nodes at all radii,
and their shapes are given in terms of the inclination angle $\theta$
relative to the plane defined by the total angular momentum vector. In
all cases there is a middle section of the disk which lies
approximately in this plane, whereas the inner and outer parts warp
away from this plane in opposite directions. For the most strongly
warped stable solution in Fig.~\ref{fig:varpdtm00018} the inner warp
is by $\sim 180\deg$ and the outer warp by $\sim 120\deg$. This is
obtained for the lowest stable pattern speed, in accordance with the
balance between gravitational and Coriolis torques [see
Fig.~\ref{fig1:mtorque} and equation (\ref{eq:nineteen})]: the torques
are weakest for the large inclinations. This can be seen already in
the two-ring problem [see equations (\ref{eq:twenty}) and
(\ref{eq:twentytwo})]. The least strongly warped disk solution in this
example has inner and outer warps $\sim$25-30$\deg$.

\begin{figure} $ \left. \right. $
     \includegraphics{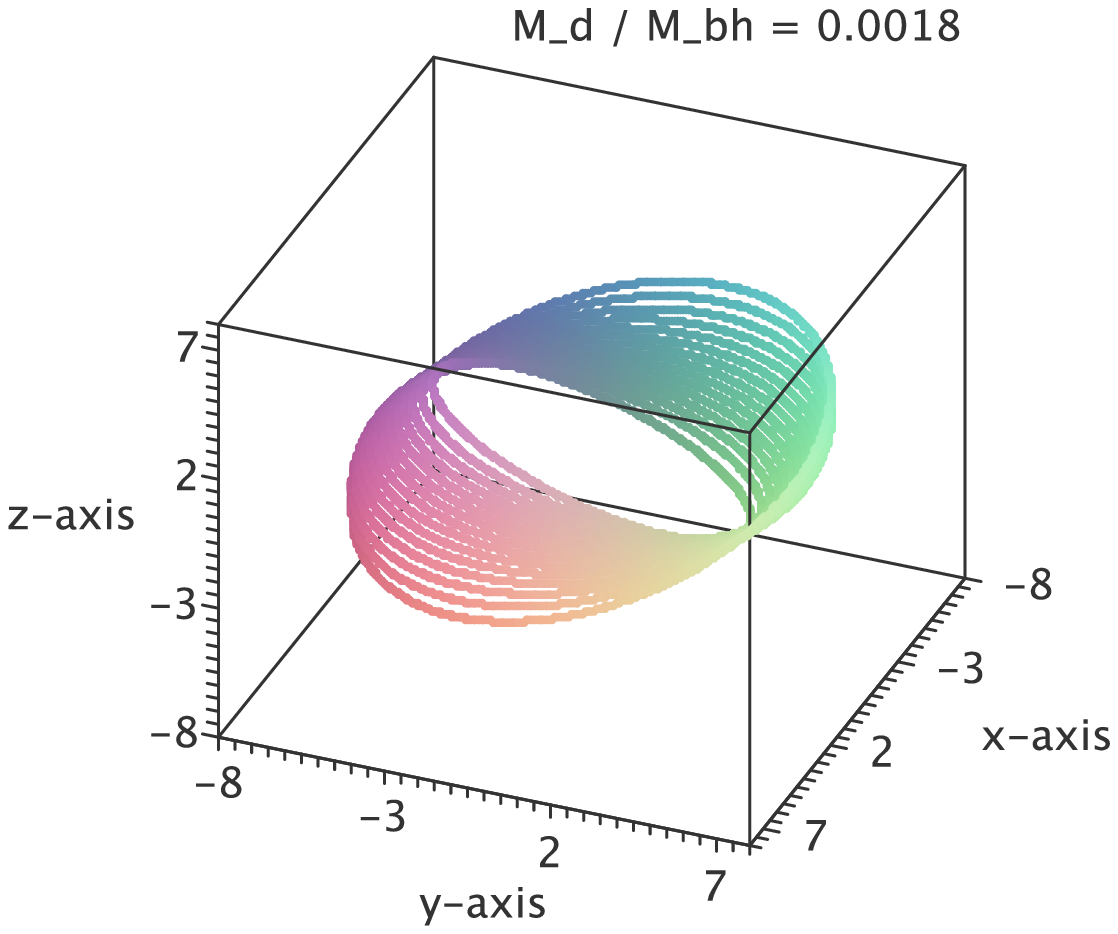}
     \includegraphics{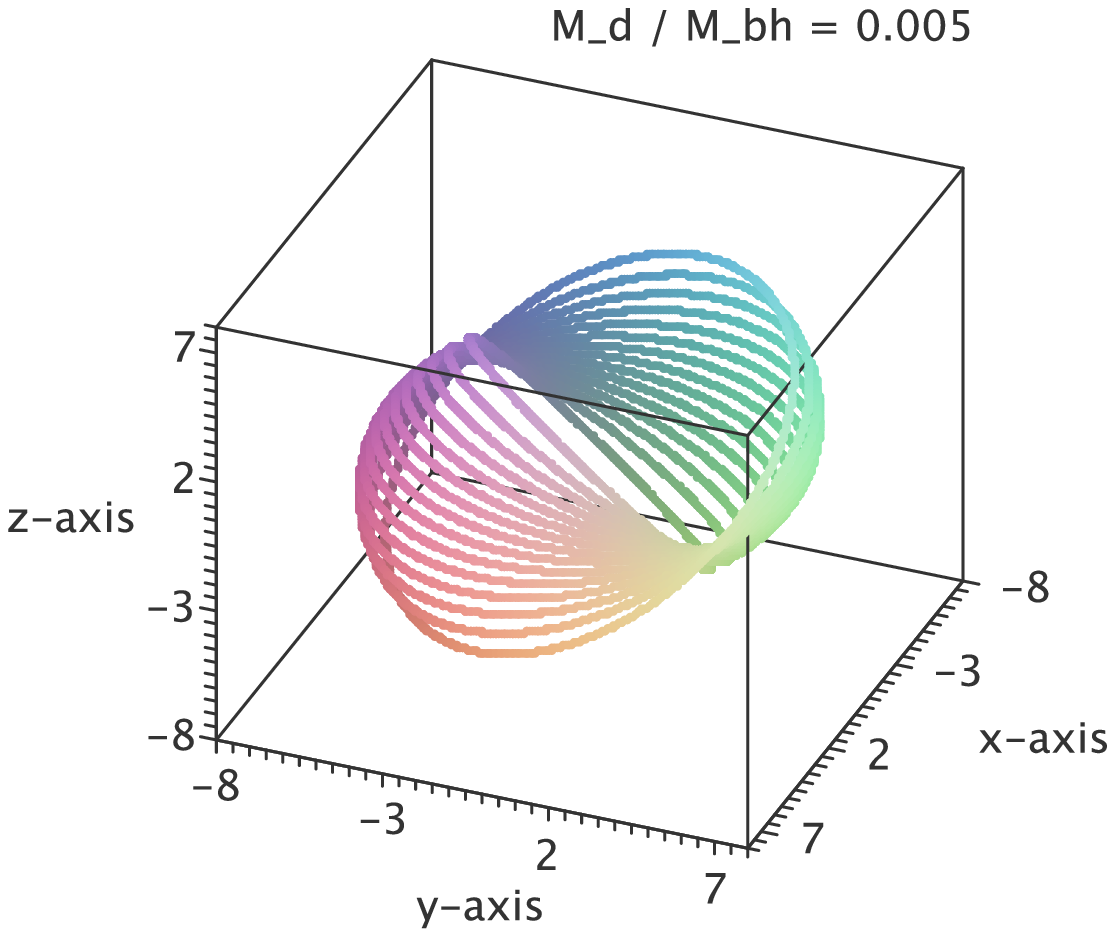}
     \includegraphics{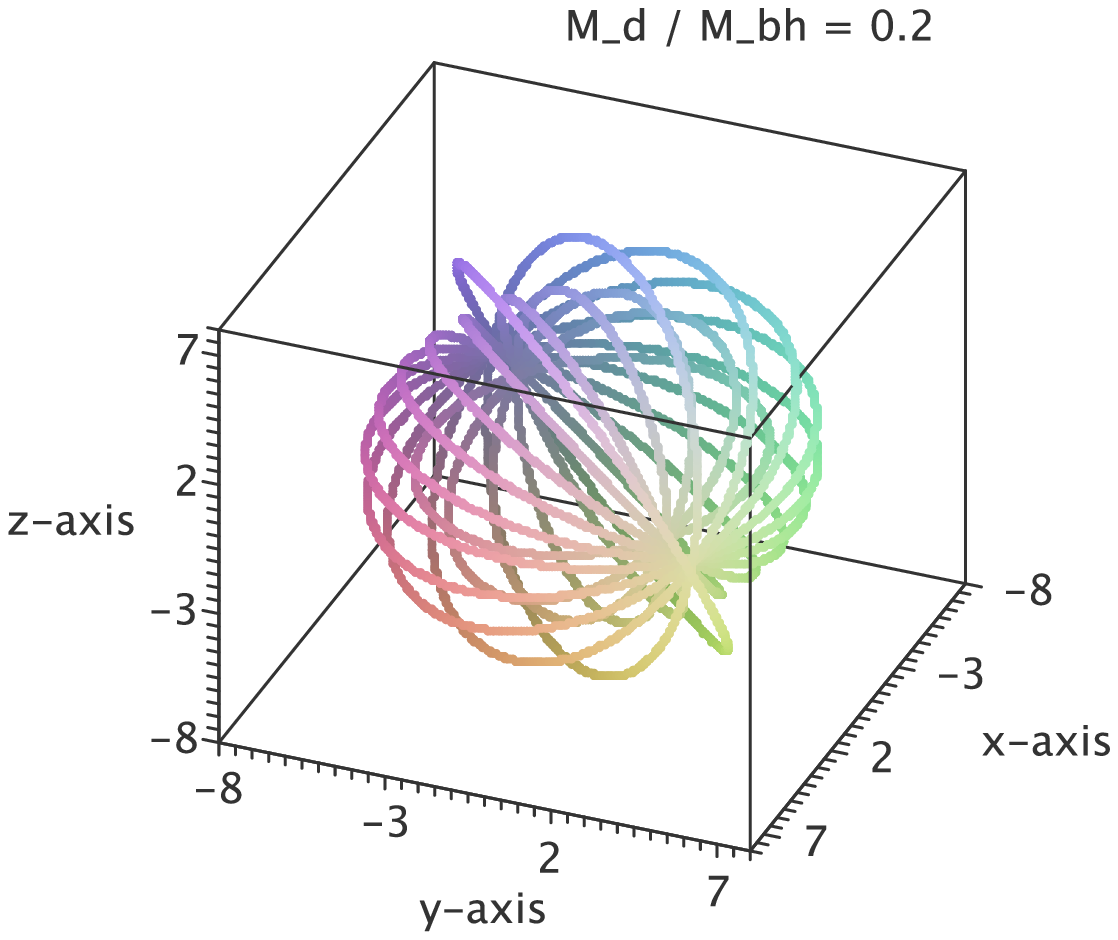}
     \vspace{21cm}
     \caption{Three-dimensional views of several 15-ring disks similar
       to those shown in Fig.~\ref{fig4:evalcsd}, with mass ratios
       given on the plots. The degree of warping increases with the
       disk mass fraction until (for disk mass fraction greater than
       $~2$ percent) the central black hole is completely hidden
       behind the warped disk.}
\label{fig:illust3d}
\end{figure}

The variation of the maximum inner and outer warp angles with disk
mass fraction is shown in Figure \ref{fig:fixphdt} for fixed
precession frequency and radial extent of the disk. The curves with
$\theta>0$ represent the outermost ring inclinations, and those with
$\theta<0$ show the innermost ring inclinations for different
precession speeds. As we have already seen in
Figs.~\ref{fig4:evalcsd}, \ref{fig:varpdtm00018}, these inclinations
increase with increasing disk mass fraction and with decreasing
precession speed.  Fig.~\ref{fig:fixphdt} also illustrates the scaling
relation of equation~(\ref{eq:twentyeight}).  Up to numerical errors,
the different curves can be scaled on top of each other.

Some three-dimensional illustrations of warped disks from this family
are shown in Figure \ref{fig:illust3d}. { From top to bottom, these
  plots shows warped disks with increasing amplitude of the warp, such
  that the disk in the bottom panel of Fig.~\ref{fig:illust3d}
  completely encloses the central black hole.}

\subsection{Comparison with Linear Theory Solutions}

\begin{figure}
\begin{centering}
\includegraphics[width=8cm]{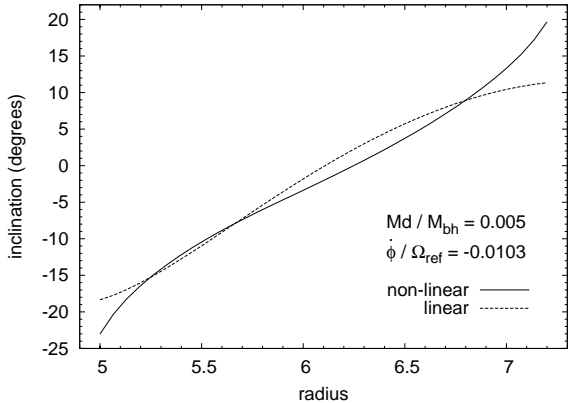}
\caption{Warp shapes for a disk with $M_d/M_{bh}=0.005$ as obtained in the 
linear (dashed line), and nonlinear (solid line) torques regimes, { for 
the same precession speed. The linear mode is scaled to the maximal
amplitude for which the linear approximation to the gravitational
torques can be used. The non-linear disk shape is determined for the
precession frequency given by the eigenvalue of the linear mode;
it is unstable. Stable non-linear warps for this mass configuration
have lower precession speeds and are more strongly warped than the
solution shown.} }
\label{fig:linvsnonlin}
\end{centering}
\end{figure}

Previous work on warped and twisted disks around black holes has often
made use of the linear approximation, in which the inclination angles of all
parts of the warped disk are assumed to be small. It is therefore interesting
to briefly consider the linear limit of our analysis above.

For small warping angles $\alpha_{ij}$, the self-gravity torques simplify
considerably. Because the leading term in equation~(\ref{eq:thirteena}) is 
already $O(\alpha_{ij})$, only
the $O(1)$ part of the $I_{ij}$ terms in this equation
need to be included, while the next order, $O(\alpha_{ij}^2)$,
can be neglected. Thus in computing $I_{ij}$ the $k^2$ term for ring $j$ 
in equation~(\ref{eq:ten}) can be approximated as:
\begin{equation}
k_{ij}^{2}\approx \frac{4 r_i r_j}{(r_i+r_j)^2}
\label{eq:thirtyfive}
\end{equation}
and the $I_{ij}$ term can be integrated to give
\begin{equation}
I_{ij} \approx  \frac{1}{\pi} 
          \left[\frac{E(k_{ij})(1-k_{ij}^{2}/2)}{1-k_{ij}^2} - K(k_{ij})\right]
                \frac{(1-k_{ij}^2/2)^{3/2}}{k_{ij}^2}.
\end{equation}
The mutual torque
on ring $i$ from ring $j$ becomes, to first order in $\alpha_{ij}$,
\begin{equation}
\frac{\partial V_{ij}}{\partial \theta_{i}} = 
                    \frac{\partial V_{ij}}{\partial \alpha_{ij}} 
                    \frac{\partial \alpha_{ij}}{\partial \theta_{i}} 
						\approx
      		    \frac{2 G m_{i}m_{j}r_{i}r_{j}I_{ij} \alpha_{ij}}
                            {(r_{i}+r_{j})^{3/2}}
                    \frac{\partial \alpha_{ij}}{\partial \theta_{i}}.
\label{eq:thirtyeight}
\end{equation}
{ For a precessing equilibrium when $\phi_i=\phi_j$,  
$\alpha_{ij}=\theta_i-\theta_j$, the equation $\dot{p}_{\theta_i}=0$ 
becomes to $O(\theta)$:
\begin{equation}
\dot{p_{\theta_i}} = \dot{\phi}^2 \frac{m_i r_i^2}{2}\theta_i-\dot{\phi}p_{\psi_i}
\theta_i -\sum_{j=1}^n 
\frac{2 G m_i m_j r_i r_j I_{ij}  (\theta_i-\theta_j)}
{(r_i^2+r_j^2)^{3/2}} =0.
\label{eq:ptdt}
\end{equation}
Equation (\ref{eq:ptdt}) is a quadratic eigenvalue problem for the
precession frequencies.  When linearized, it transforms into a
generalized eigenvalue problem of dimensions $2n \times 2n$. We use
the NAG routine F02BJF to find the eigenvalues and eigenvectors of
equation (\ref{eq:ptdt}).

The $2n$ eigenvalues constitute two distinct families in the frequency
spectrum of the disk; the fast prograde, and the slow retrograde
frequencies. When sorted in decreasing order, the first retrograde
frequency has a value of zero, and the associated eigenvector
represents a tilt of the whole disk by the same angle,
i.e. $\theta_i=\rm constant$.  The next eigenvalue corresponds to the
warps we have discussed so far where the disk has one radial node
\citep[modified tilt mode,][]{hunter69,sparke84,sparke88}. In the
following, we restrict our discussion to linear warp shapes of this
kind.

In Figure \ref{fig:linvsnonlin} we show the modified tilt mode in
linear theory of a disk with $M_d/M_{bh}=0.005$, $r_{\rm in}=5$ and
$r_{\rm out}=7.2$ for 40 rings. This is obtained by solving equation
(\ref{eq:ptdt}) and is shown with the dashed line. We note that in
linear theory the warp shape can be arbitrarily scaled as long as the
local gradient of the tilt satisfies $d \theta/ dr < 1.2 /r$ (see
Section (\ref{sec:compoto}); if this condition is violated, the linear
approximation to the self-gravity torques breaks down.  Therefore in
Fig.~\ref{fig:linvsnonlin} the linear mode is scaled to its maximum
possible amplitude such that the condition is everywhere satisfied.

As mentioned above, the precession frequency of this mode is the first
nontrivial eigenvalue in the retrograde family, and here it has a
value of $\dot{\phi} / \Omega_{\rm ref}=-0.0103$ when normalized to
the rotation frequency of the reference ring. For this frequency, we
then solve equation (\ref{eq:fivea}) to obtain the nonlinear warp
shape shown by the solid line in Figure~\ref{fig:linvsnonlin}. The
larger curvature of the non-linear warp near the inner and outer
boundaries of the disk, with respect to the scaled linear mode, shows
that the linear approximation overestimates the torques in these parts
of the disk.

However, the main difference between linear modes and
non-linear warps is that, for a given mass distribution of the disk
(surface density profile, inner and outer boundaries), the precession
frequency and shape of the modified tilt mode in linear theory is
uniquely determined, whereas non-linear equilibrium warp solutions may
exist for a range of precession frequencies and warp shapes or, e.g.,
for extended disks, may not exist at all. For the case shown in
Fig.~\ref{fig:linvsnonlin}, non-linear warped equilibria are found
for precession frequencies in the range
$\dot{\phi}/\Omega_{\rm ref}=-8.87\times 10^{-5} \rightarrow
-1.16\times10^{-2}$ and are stable in the range
$\dot{\phi}/\Omega_{\rm ref}=-9.24\times 10^{-5} \rightarrow
-9.63\times10^{-3}$. The particular non-linear warp shape obtained for
the frequency of the linear mode and shown in
Fig.~\ref{fig:linvsnonlin} is unstable. 

Alternatively, the warp shape may be parametrized by the inclination
of the outermost ring, say, $\theta_n$. Linear theory warps can in the
previous example be considered valid up to $\theta_n\simeq 10^\circ$,
and have all the same precession frequency. Non-linear warp modes are
found in the range $\theta_n = 19.1^\circ \rightarrow 130^\circ$, and
are stable in the range $\theta_n = 20.9^\circ \rightarrow
129^\circ$. They are disjunct in $\theta_n$ from the linear modes, and
their precession speed decreases with $\theta_n$ according to the
balance of gravitational and Coriolis torques.

}


\subsection{Dependence on Surface Density Profile and Radial Extent}

\begin{figure}
\includegraphics[angle=0,width=8cm]{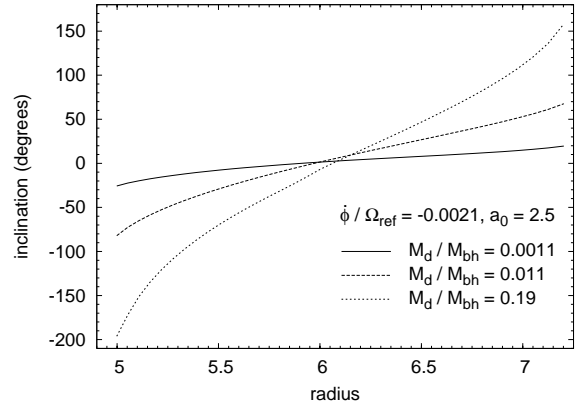}
\centering
\caption{Inclination of a 35-ring-disk at different radii as in 
Figure~\ref{fig4:evalcsd}, but for an exponential surface density
profile with scale length $r_d=2.5$.  The precession speed is
identical to that in Fig.~\ref{fig4:evalcsd}.  The upper and lower
curves correspond to the limiting mass ratios for disk stability; see
Section \ref{sec:anstab}.  }
\label{fig:evalexpo}
\end{figure}

The warped disks presented in Figs.~\ref{fig4:evalcsd},
\ref{fig:fixphdt} have constant surface density. For comparison,
Figure~\ref{fig:evalexpo} shows the warping of an exponential disk,
with surface density $\Sigma(r)=\Sigma_{0} \exp(-r/r_d)$ where
$\Sigma_{0}$ denotes the central density and $r_d$ is the scale
length, chosen to be 2.5 units in this example. The other parameters
(relative ring radii, precession speed) are identical to those used in
Fig.~\ref{fig4:evalcsd}. The basic warp shapes are similar as for
constant surface density, but the maximum outer warp angles are
slightly larger. The range of stable disk masses is also comparable to
that for the constant surface density disk (for the same precession
speed); see the curves showing the boundaries of stability in
Figs.~\ref{fig4:evalcsd}, \ref{fig:evalexpo}. 

\begin{figure}
\centering
\includegraphics[angle=0,width=8cm]{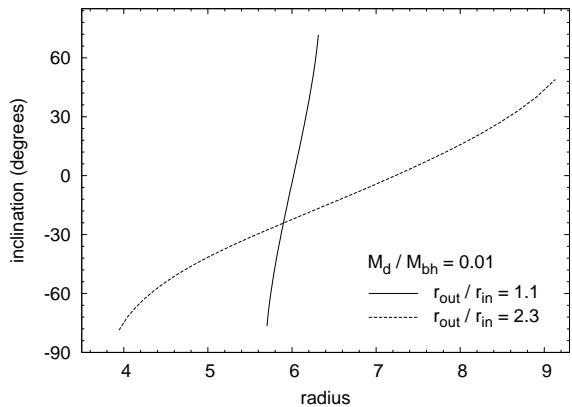}
\caption{Stable warped equilibria for 35-ring-disks with
constant surface density but varying radial extent, and with
$M_{d}=0.01 M_{bh}$.  The solid and {{dashed}} lines demarcate the
boundaries of the region of stable solutions. }
\label{fig:evalvarbetam001}
\end{figure}

\begin{figure}
\centering
\includegraphics[angle=0,width=8cm]{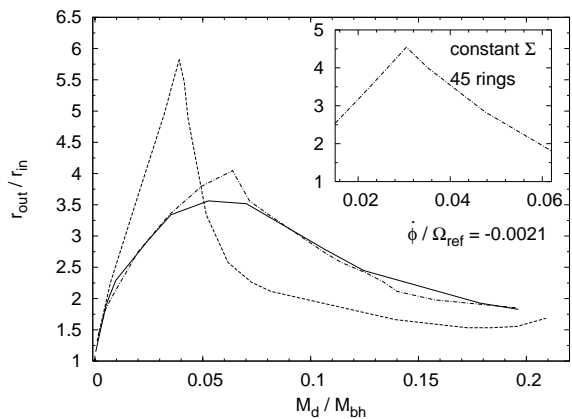}
\caption{Maximum radial extent of stable warped disk equilibria
  $r_{\rm out}/r_{\rm in}$, as a function of disk-to-black hole mass
  ratio. The dashed line represents the result for a disk of 15 rings
  with constant surface density, the solid line is for an exponential
  disk with scale-length $r_d=5.2$, and the dot-dashed line is for an
  exponential disk with scale-length $r_d=2.5$. The inset shows
  results obtained for the constant surface density disks when instead
  approximated by 45 rings, near the mass ratio with maximum $r_{\rm
    out}/r_{\rm in}$. Typical ratios are $r_{\rm out}/r_{\rm
    in}=$2-4.}
\label{fig:massvsbeta_com}
\end{figure}

Because the condition for a warped equilibrium is that the Coriolis
and gravitational torques balance, clearly not only the mass fraction
and mass distribution, but also the radial extent of the disk must be
important for determining the warp shape and its stability.  To
investigate this we compiled a set of precessing equilibria with
varying radius scaling factor $\kappa$, as follows (see also
Section~\ref{sec:nringapprox}).  After fixing the radius of the middle
ring of the disk, $r_{\rm mid}$, we determine the remaining ring radii
such that
\begin{equation}
r_i= r_{\rm mid} \kappa^j
\begin{cases}
  j=i-\frac{n+1}{2}      & \text{$n$ odd} \\
  j=i-\frac{n}{2}        & \text{$n$ even}
\end{cases} 
   \qquad i=1,2,\ldots,n
\label{eq:fourtytwo}
\end{equation}
where $n$ is the number of rings. For illustration, we consider a
family of disk models with the same disk-to-black hole mass fraction
$M_d=0.01M_{bh}$, each with its own constant surface density given by
$M_d$ and $\kappa$.  All disks are made of $n=35$ rings, and the
middle ring radius is set to $r_{\rm mid}=6$ units.

Figure \ref{fig:evalvarbetam001} shows precessing equilibria for such
disks for different $\kappa$. The upper and lower curves show the two
disk shapes that bound the stable range of solutions in terms of the
$\kappa$-factor.  In the case where the rings have minimum possible
separation from each other, the inner ring has a radius of $r_{1}=5.7$
units, and the outer ring has $r_{n}=6.3$ units. On the other hand,
for the most extended stable disk in this family, $r_{1}=3.9$, and
$r_{n}=9.1$.  When the extent of the disk is increased, a slight
decrease in the warping is observed in
Fig.~\ref{fig:evalvarbetam001}. This is due to the fact that the torque
from a ring of constant mass decreases with distance to the ring,
cf.\ equation~(\ref{eq:thirteena}).  

Figure \ref{fig:massvsbeta_com} shows the radial extent of the disk
$r_{\rm out}/r_{\rm in}$ for which stable warped equilibria can be
found, for different surface density profiles and as a function of
disk-to-black hole mass ratio.  The most important result of these
calculations is that stable non-linear warps can be maintained only
for disks with inner and outer boundaries, for which $r_{\rm
  out}/r_{\rm in}\simeq$ 2-4. This is reminiscent of the result 
of \citet{hunter69} that in linear theory only truncated disks 
permit long-lasting bending modes.

\subsection{Time-Evolution of Ring Systems}
\label{sec:timevol} 
In this section we consider the explicit time-evolution of a
precessing system of self-gravitating rings in a massive black hole
potential. By integrating the equations of motion,
equations~(\ref{eq:three})-(\ref{eq:six}), starting from initial
conditions corresponding to one of the precessing disk solutions found
earlier, we can check the stability of this solution directly and
compare with the linear stability analysis.

In these integrations, we use disks of 20 equal mass rings,
equally spaced in radius.  The ratio of the outermost ring radius to
that of the innermost ring is $1.44$.  The initial $\theta_{i}$ are
obtained from precessing equilibrium solutions; all rings have the
same line-of-nodes, i.e., the same initial $\phi_i$. The equilibrium
precession speed is given by $\dot{\phi}/ \Omega_{\rm ref}=-0.0021$.

In the following figures, symbols starting from the outer circle show
the variation of inclination $\theta$ with ring radius, where the ring
radii are shown as distances from the center of the plot, with scale
shown on the lower right. The symbols starting from the inner circle
show how the azimuthal angle $\phi$ changes with the ring radius; for
this part of the plot, the ring radii are scaled down to the half of
their values to make the figure more easily readable.  The elapsed
time of the integration is shown on top of the figures, in terms of
the number of orbital periods at the position of the outermost ring
$n$ where $\dot{\phi}/ \Omega_{n}=-0.0027$.

\begin{figure} $ \left. \right. $
     \includegraphics{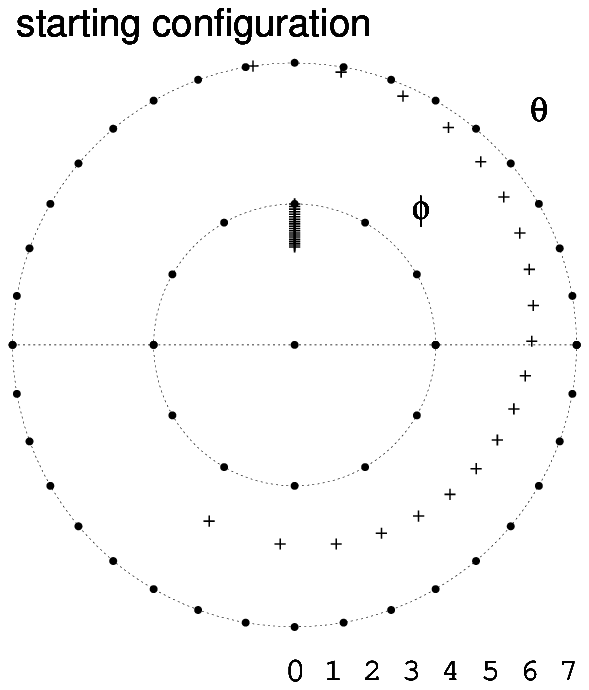}
     \includegraphics{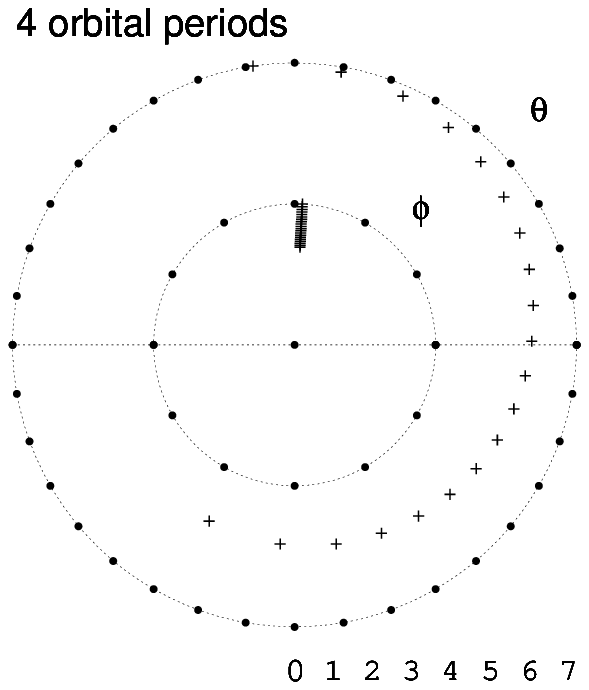}
     \includegraphics{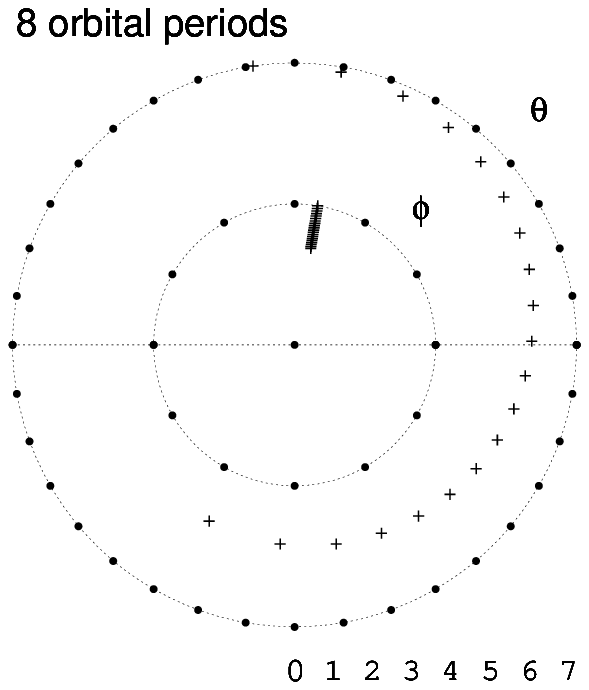}
     \includegraphics{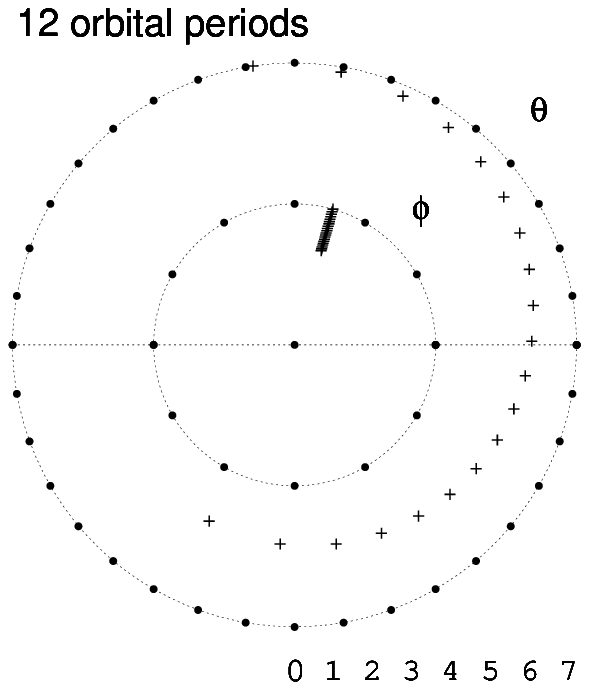}
\vspace{10cm}
\caption{Time-evolution of a disk of 20 rings with total mass $M_{d}= 0.05 M_{bh}$. 
The disk is stable.}
\label{fig:20rm005}
\end{figure}

\begin{figure} $ \left. \right. $
     \includegraphics{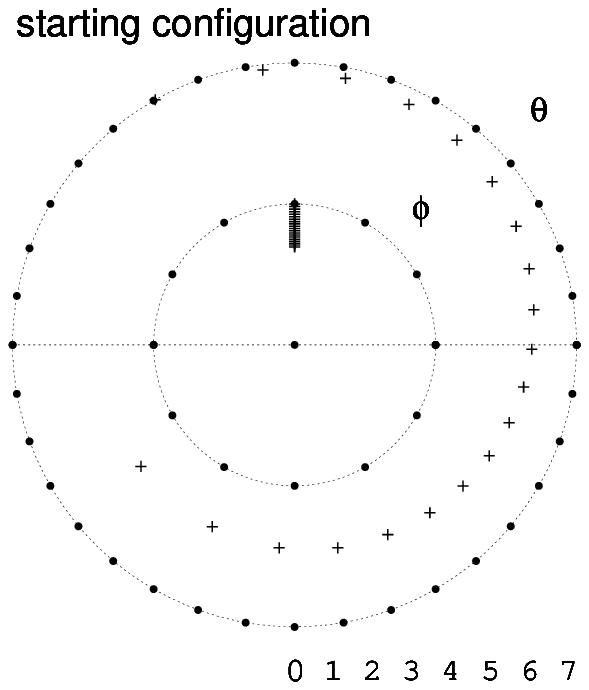}
     \includegraphics{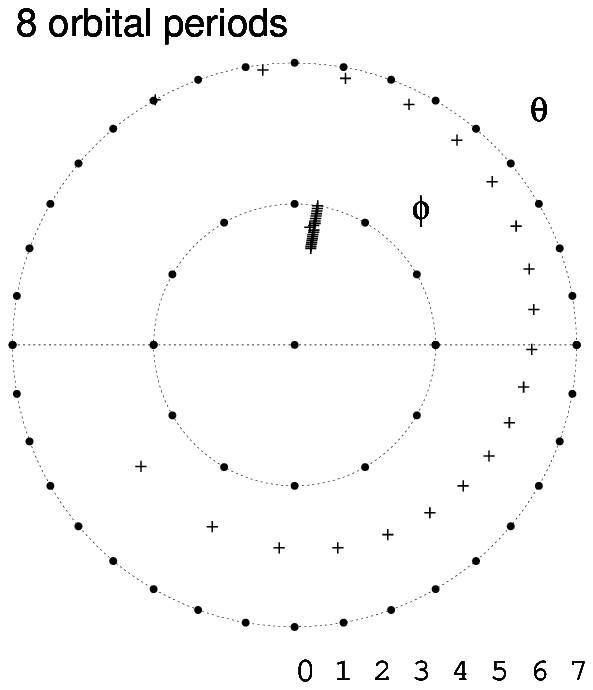}
     \includegraphics{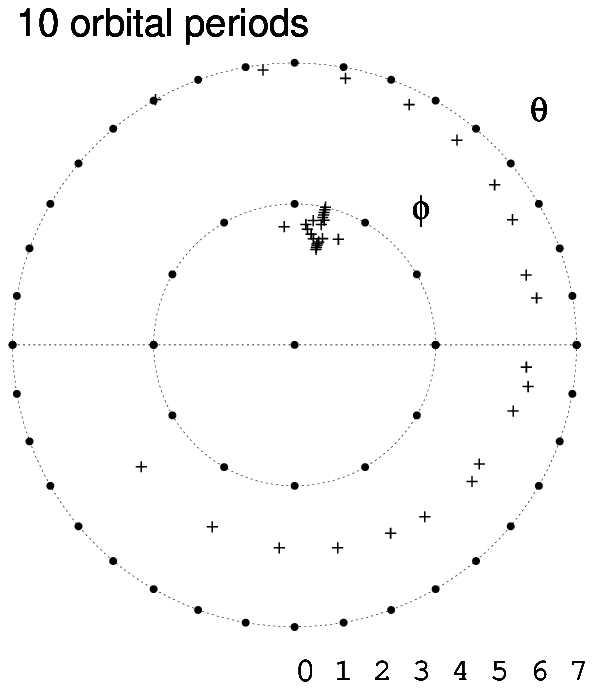}
     \includegraphics{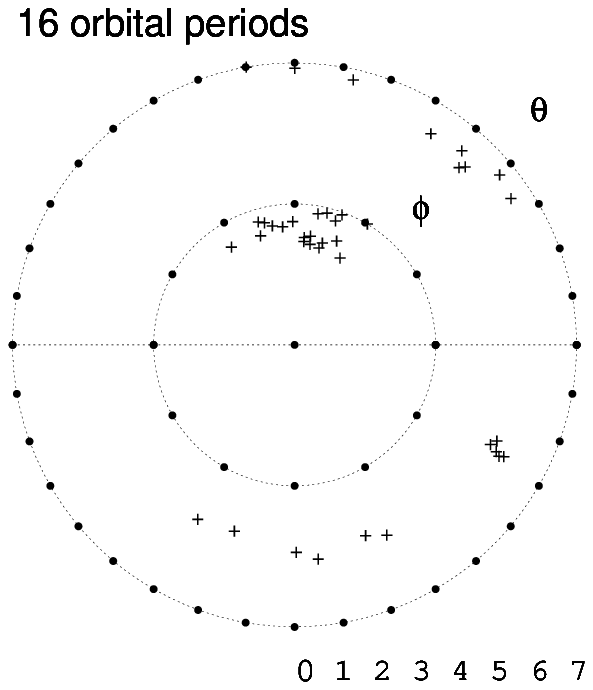}
     \vspace{10cm}
\caption{Time evolution of a disk of 20 rings with total mass $M_{d}=0.1 M_{bh}$. The disk
precesses as a unit for nearly 8 orbital periods, but then it starts to
break up, so is unstable.}
\label{fig:20rm01}
\end{figure}

Figure \ref{fig:20rm005} shows the time evolution of a disk of 20
rings with $M_{d}=0.05 M_{bh}$. The disk stays in equilibrium for $12$
orbital periods, consistent with its linear stability.  Figure
\ref{fig:20rm01} shows the evolution of disk of 20 rings with $M_{d}=
0.1 M_{bh} $. This disk precesses as a unit for 8 orbital periods, but
then it starts to break into parts, hence the disk is unstable, as also
predicted by linear stability analysis.

\begin{figure} $ \left. \right. $
     \includegraphics{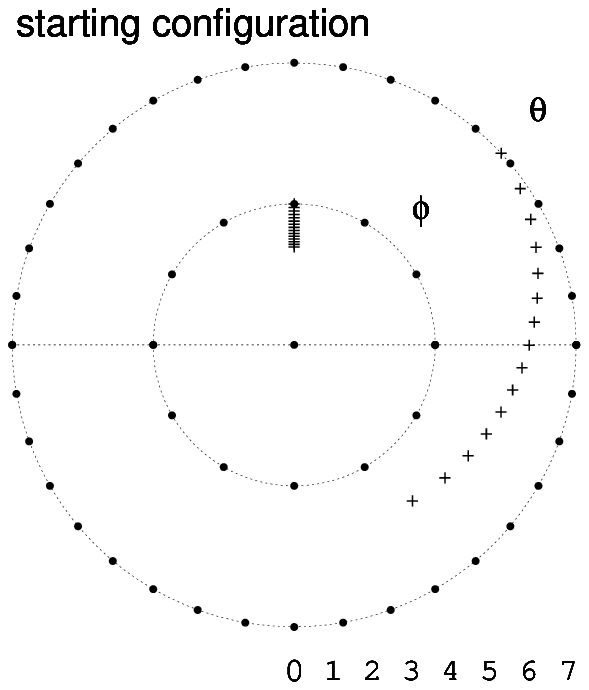}
     \includegraphics{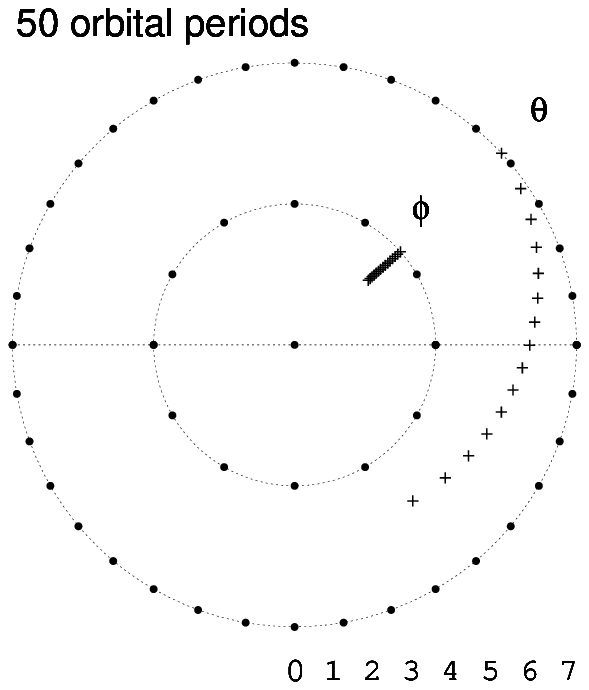}
\vspace{4.5cm}
\caption{Time-evolution of a disk of 15 rings with mass fraction $M_{d}/M_{bh}=0.005$. 
The disk is followed for $50$ orbital periods and is stable.}
\label{fig:fig23}
\end{figure}

\begin{figure} $ \left. \right. $
     \includegraphics{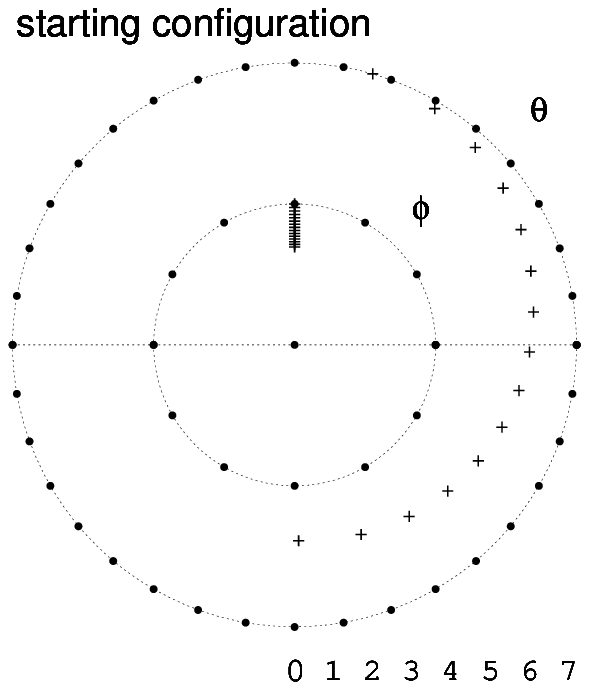}
     \includegraphics{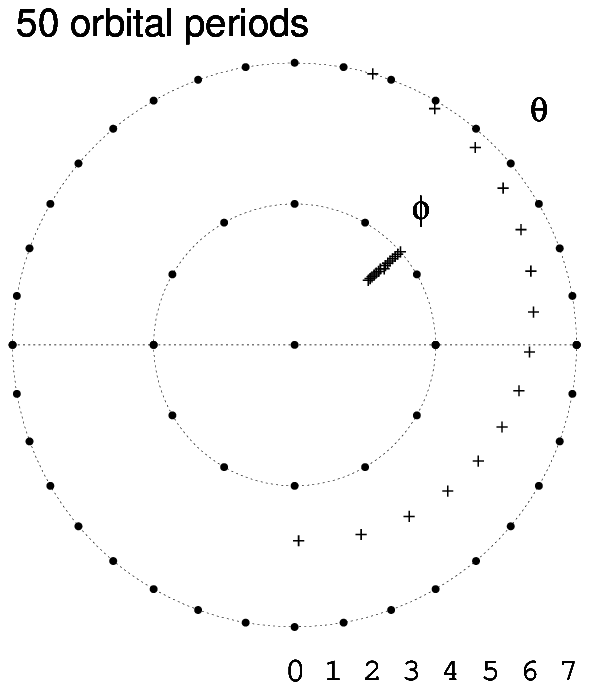}
\vspace{4.5cm}
\caption{Time-evolution of a disk of 15 rings with mass fraction of $M_{d}/M_{bh}=0.02$. 
The evolution is followed for $50$ orbital periods and the disk is stable.}
\label{fig:fig24}
\end{figure}

To strengthen the agreement between linear stability and time
evolution results, we integrate two of these ring systems for
longer. Figures \ref{fig:fig23} and \ref{fig:fig24} show the time
evolution of two disks with masses $M_{d}=0.005 M_{bh}$ and $M_d =
0.02M_{bh}$, with 15 logarithmically spaced rings according to
equation~(\ref{eq:fourtytwo}), over 50 orbital periods.  In both cases
the disks are stable, as expected from the linear stability analysis.

\section{Discussion}

\subsection{Theoretical Issues}

In this work we have considered warped disks around black holes for
which the only acting force is gravity and the disk is approximated as
a nested sequence of circular rings. We have focussed on non-linearly
warped, steadily precessing disk configurations, contrary to most
previous work in which small amplitude warps were considered, often of
a transient nature. We have found that stable, steadily precessing,
highly warped disks can be constructed, albeit only over a limited
radial range, such that the typical ratio of the outer to the inner
boundary radius is $\sim2$-4. 

In one illustrative case, we have compared with a linear theory warped
disk. For a given disk mass configuration, the precession frequency of
the linear, modified tilt mode is given as an eigenvalue, and the
shape can be scaled up to the amplitude where the validity of the
linear approximation to the gravitational torques breaks down. The
corresponding non-linear warp with the same precession frequency is
unstable. Stable non-linear warps for the same mass configuration
exist for a disjunct range of precession speeds which are all slower
than that of the linear mode. Their warp angles increase with
decreasing precession speed, and the non-linear solutions are more
strongly warped than the linear mode at the maximum scaling.

These warped disks obey a scaling relation in the sense that (i) they
can be scaled to an arbitrary radius $r$, provided the precession
speed is scaled to the circular frequency $\Omega(r)$, and (ii) they
can be scaled in mass, provided the ratio of precession frequency to
$\Omega(r)$ keeps in line with the ratio of disk mass to black hole
mass.

\begin{figure}
\centering
\includegraphics[angle=0,width=8.5cm]{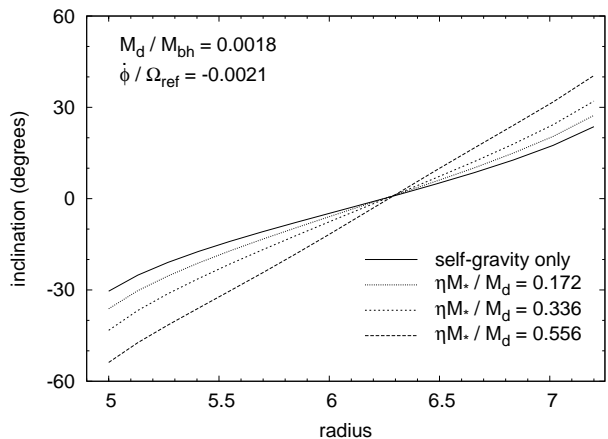}
\caption{Inclination of a disk of constant surface density at different
radii, under the self-gravity torques and the quadrupole torques of
a surrounding star cluster. This is parametrized by the ratio 
$\eta M_*/M_d$, where $\eta$ measures the flattening of the potential
(equation (2) of \citealt{sparke86}) and $M_*/M_d$ is the ratio of the
mass of the star cluster to the mass of the disk, inside the outermost
ring radius.}
\label{fig:quad}
\end{figure}

In constructing these solutions, we have neglected the background
potential generated by the surrounding nuclear star cluster, whose
quadruple moment will often be important on scales of $\sim 0.1$ pc.
Figure \ref{fig:quad} shows steadily precessing warped disk
inclinations for one case including the background potential. For the
parameters chosen, the solutions are qualitatively similar to those
discussed earlier.

Stability was tested with respect to perturbations of the ring
parameters, that is, the orbits of gas and stars were assumed to
remain circular.  We did not investigate instabilities by which the
disk would become eccentric or lop-sided.  Answering the question
whether such instabilities are relevant for the warped disks
considered here requires different techniques and must remain for
future work \citep[see, e.g.,][]{touma02}.

Neglecting gas pressure and viscosity for our warped disk solutions is
justified if the disks are cold and the viscous time-scale is much
longer than the precession time-scale. \cite{pringle92} has devised a
system of equations for the evolution of the surface density and local
angular momentum vector of a non-linearly warped, viscous disk. A
logical next step is to add the gravitational torques to these
equations and study the evolution of viscous, self-gravitating,
non-linearly warped disks; this is work in progress.

\subsection{Origin of Warped Disks}

An important question is whether, and if so how, the warped nuclear
disks we have considered can be set up in nature. { Infall of gas
  clouds on inclined orbits has been discussed in the context of
  observations of the Galactic center (see next section) as a possible
  model for generating a warped disk in the central parsec
  \citep{hobbs+nayakshin09}.  If the potential of the nuclear star
  cluster is important, accretion of gas onto a plane inclined
  relative to its principal plane may lead to a warped disk. The
  combined quadrupole moment of the gas disk itself and of the
  background cluster potential would cause the orbits to precess and
  the disk to become warped.  In both cases, the accreting gaseous
  material with misaligned angular momenta will not directly end up in
  a warped disk with the right density structure for steady state
  precession. However, the disk may settle into a warp mode if the
  energy associated with the transient response can be transported
  outwards by bending waves \citep{toomre83,hofner94}, or in the case
  of gaseous disks, if it can be dissipated \citep[see discussion
  in][]{papaloizou98}; this remains to be investigated. }

\citet{caproni06b} discuss four warping mechanisms for extragalactic
accretion disks: tidal, irradiative, magnetic, and Bardeen-Petterson.
If a planar disk has become warped by the radiation pressure
instability discussed by \citet{petterson77,pringle96} or through
magnetic instabilities \citep{lai03}, the gravitational torques might
start to dominate once the source of the initial warping disappears.
{ Highly warped disks have been reported before by
  \citet{pringle97} in the context of the radiation pressure
  instability. We have done some simple time evolution calculations to
  show that initially highly warped disks often do not dissolve
  through self-gravity precession; the torques then cause wobbling but
  not break-up of the disk. }  The role of self-gravity in
such models would be to ensure the long term { persistence} of the
warp. Future work along the lines discussed at the end of the last
subsection may be able to clarify whether this is feasible.

\subsection{Warped Disks in Galactic Nuclei}

Warped disks around central black holes have been inferred through
observations of water maser emission in several nearby active galaxies
such as NGC 4258, NGC 1068, and the Circinus galaxy.  The maser disks
in these galaxies extend radially between $0.16-0.28$ pc
\citep{herrnstein99}, $0.65-0.11$ pc \citep{greenhill97}, and
$0.11-0.4$ pc \citep{green03}, respectively. The most widely studied
of these maser disks is in NGC 4258, where from the near-Keplerian
rotation curve of the high-velocity masers the black hole mass is
deduced to be $3.8\times10^7\msun$, and the dynamical upper limit to
the mass of the disk is $< 10^6\msun$ \citep{herrnstein+05}.
Stationary, power-law accretion disk models constrained by theory and
observations have mass fractions $10^{-4}-10^{-3}$ of the central
black hole, in which case the gravitational and viscous torques are
comparable \citep{caproni07,martin08}. Several explanations have been
suggested for the observed warp in the disk \citep{caproni06b}. In one
model the warp is caused by a binary companion orbiting outside the
disk \citep{papaloizou98}; this would need a mass comparable to that
of the disk.  A second possibility is radiation pressure from the
central source \citep{pringle96, maloney96}, but \citet{caproni06b}
analysing several AGN disks find that these are stable against
radiation warping. The most favoured explanation for the warp is the
Bardeen-Petterson effect \citep{caproni07, martin08} but to reach a
steady state the disk must be very long-lived.  Gravitational torques
have so far been { mostly} neglected; our results suggest that it
may be worthwhile to consider models including both the gravity from
the disk and possibly the quadrupole moment of the stellar cusp.

In the Galactic center, near-infrared observations have identified one or
possibly two disks of young stars at a distance of $\sim 0.04 \rm \ to
\ 0.4$ pc from the central black hole SgrA*
\citep{genzel03,paumard06,lu+09,bartko08}.  These stellar disks are
highly inclined both with respect to the Galactic plane, and with
respect to each other. The total mass in the disks, as inferred from
stellar number counts, is around $10^4 \rm \ M_{\odot}$
\citep{paumard06}. This is a non-negligible fraction of the mass of
SgrA*, $M_{\rm bh}\sim 4 \times 10^6 \rm \ M_{\odot}$
\citep{genzel00,ghez05}. The recent analysis of \citet{bartko08}
  shows that the clockwise rotating disk is warped, with angular
  momentum direction slewing over $\sim 60^{\circ}$ from the inner to
  the outer stars.  We consider the precession of the warped disk in
  the Galactic center elsewhere.

Warped disks could also have important implications for the
unification of AGN \citep{phinney89}. The unification theories rely on
the obscuration along some lines-of-sight of the radiation from the
central source by intervening matter. While this obscuring matter is
usually depicted as a doughnut-like torus, an alternative possibility
is that it could have the shape of a flared or warped disk.  { The
  highly warped solutions discussed above in principle provide the
  geometry to obscure the central engine from most lines-of-sight.}
The obscuring medium required for these unification scenarios must be
clumpy \citep{nenkova02}, perhaps suggesting fragmentation of the disk
\citep{goodman03}.  \cite{nayakshin05} studied the evolution of a
highly inclined warped disk, where he showed that the disk indeed can
conceal the central object for most of its lifetime. In the nonlinear
regime, warped disks can obscure a significant part of the solid angle
of the source (see Figure~\ref{fig:illust3d} in Section
\ref{sec:results} above).  Recently, \cite{wu08} showed that because
the outer parts of a warped disk receive a larger fraction of the
central emission, the line ratios of the reprocessed Balmer emission
lines can be successfully predicted by a warped disk model.

\section{Summary and Conclusions}
\label{sec:summ}

In this paper we have investigated { non-linearly} warped
disk solutions around black holes for which the only acting force is
gravity. We used a simple model in which the disk is approximated as a
nested sequence of circular rings. We have shown that with these
approximations stable, steadily precessing, highly warped disks can be
constructed.

{ These disks have a common line-of-nodes for all rings. In all
  cases there is a middle section of the disk which lies approximately
  in this plane, whereas the inner and outer parts warp away from this
  plane in opposite directions. The warp angles of these solutions can
  be very large, up to $\sim\pm120$ deg, but they extend only over a
  limited radial range, such that the typical ratio of the outer to
  the inner boundary radius is $\sim2$-4. Such precessing equilibria
  exist for a wide range of disk-to-black hole mass ratios
  $M_d/M_{bh}$, including quite massive disks.}

The stability of these precessing disks was determined using linear
perturbation theory and, in a few cases, confirmed by numerical
integration of the equations of motion. { We found that over most
  of the parameter range investigated, the precessing equilibria are
  stable, but some are unstable.}

These disks obey a scaling relation: they can be scaled to arbitrary
radii $r$, provided the precession speed is scaled to the circular
frequency $\Omega(r)$, and they can be scaled in mass, provided the
ratio of precession frequency to $\Omega(r)$ is changed, in good
approximation, proportionally to the ratio of disk mass to black hole
mass.

{ The main result of this study is} that persistent forcing of the
disk other than by its own self-gravity is not necessarily required
for maintaining a non-linearly warped disk in a Keplerian potential.
Further work combining self-gravity with gas physics etc.\ will show
whether these self-gravitating warped disk solutions help to
understand the observed warped disks in galactic nuclei.


\bibliography{P1Disksunbold}

\begin{thebibliography}{48}
\expandafter\ifx\csname natexlab\endcsname\relax\def\natexlab#1{#1}\fi

\bibitem[{{Arnaboldi} \& {Sparke}(1994)}]{arnaboldi}
{Arnaboldi}, M. \& {Sparke}, L.~S. 1994, \aj, 107, 958

\bibitem[{{Bardeen} \& {Petterson}(1975)}]{bardeen75}
{Bardeen}, J.~M. \& {Petterson}, J.~A. 1975, \apjl, 195, L65+

\bibitem[{{Bartko} {et~al.}(2009){Bartko}, {Martins}, {Fritz}, {Genzel},
  {Levin}, {Perets}, {Paumard}, {Nayakshin}, {Gerhard}, {Alexander},
  {Dodds-Eden}, {Eisenhauer}, {Gillessen}, {Mascetti}, {Ott}, {Perrin},
  {Pfuhl}, {Reid}, {Rouan}, {Sternberg}, \& {Trippe}}]{bartko08}
{Bartko}, H., {Martins}, F., {Fritz}, T.~K., {Genzel}, R., {Levin}, Y.,
  {Perets}, H.~B., {Paumard}, T., {Nayakshin}, S., {Gerhard}, O., {Alexander},
  T., {Dodds-Eden}, K., {Eisenhauer}, F., {Gillessen}, S., {Mascetti}, L.,
  {Ott}, T., {Perrin}, G., {Pfuhl}, O., {Reid}, M.~J., {Rouan}, D.,
  {Sternberg}, A., \& {Trippe}, S. 2009, \apj, 697, 1741

\bibitem[{{Binney} {et~al.}(1998){Binney}, {Jiang}, \& {Dutta}}]{binney98}
{Binney}, J., {Jiang}, I.-G., \& {Dutta}, S. 1998, \mnras, 297, 1237

\bibitem[{{Binney} \& {Tremaine}(1987)}]{binney}
{Binney}, J. \& {Tremaine}, S. 1987, {Galactic dynamics} (Princeton, NJ,
  Princeton University Press, 1987, 747 p.)

\bibitem[{{Caproni} {et~al.}(2007){Caproni}, {Abraham}, {Livio}, \& {Mosquera
  Cuesta}}]{caproni07}
{Caproni}, A., {Abraham}, Z., {Livio}, M., \& {Mosquera Cuesta}, H.~J. 2007,
  \mnras, 379, 135

\bibitem[{{Caproni} {et~al.}(2006){Caproni}, {Livio}, {Abraham}, \& {Mosquera
  Cuesta}}]{caproni06b}
{Caproni}, A., {Livio}, M., {Abraham}, Z., \& {Mosquera Cuesta}, H.~J. 2006,
  \apj, 653, 112

\bibitem[{{Gallimore} {et~al.}(2004){Gallimore}, {Baum}, \&
  {O'Dea}}]{gallimore04}
{Gallimore}, J.~F., {Baum}, S.~A., \& {O'Dea}, C.~P. 2004, \apj, 613, 794

\bibitem[{{Genzel} {et~al.}(2000){Genzel}, {Pichon}, {Eckart}, {Gerhard}, \&
  {Ott}}]{genzel00}
{Genzel}, R., {Pichon}, C., {Eckart}, A., {Gerhard}, O.~E., \& {Ott}, T. 2000,
  \mnras, 317, 348

\bibitem[{{Genzel} {et~al.}(2003){Genzel}, {Sch{\"o}del}, {Ott}, {Eisenhauer},
  {Hofmann}, {Lehnert}, {Eckart}, {Alexander}, {Sternberg}, {Lenzen},
  {Cl{\'e}net}, {Lacombe}, {Rouan}, {Renzini}, \& {Tacconi-Garman}}]{genzel03}
{Genzel}, R., {Sch{\"o}del}, R., {Ott}, T., {Eisenhauer}, F., {Hofmann}, R.,
  {Lehnert}, M., {Eckart}, A., {Alexander}, T., {Sternberg}, A., {Lenzen}, R.,
  {Cl{\'e}net}, Y., {Lacombe}, F., {Rouan}, D., {Renzini}, A., \&
  {Tacconi-Garman}, L.~E. 2003, \apj, 594, 812

\bibitem[{{Ghez} {et~al.}(2005){Ghez}, {Salim}, {Hornstein}, {Tanner}, {Lu},
  {Morris}, {Becklin}, \& {Duch{\^e}ne}}]{ghez05}
{Ghez}, A.~M., {Salim}, S., {Hornstein}, S.~D., {Tanner}, A., {Lu}, J.~R.,
  {Morris}, M., {Becklin}, E.~E., \& {Duch{\^e}ne}, G. 2005, \apj, 620, 744

\bibitem[{{Goldreich}(1966)}]{goldreich}
{Goldreich}, P. 1966, Reviews of Geophysics, 4, 411

\bibitem[{{Goldstein} {et~al.}(2002){Goldstein}, {Poole}, \&
  {Safko}}]{goldstein}
{Goldstein}, H., {Poole}, C., \& {Safko}, J. 2002, {Classical mechanics}
  (Classical mechanics (3rd ed.) by H.~Goldstein, C.~Poolo, and J.~Safko.~San
  Francisco: Addison-Wesley, 2002.)

\bibitem[{{Goodman}(2003)}]{goodman03}
{Goodman}, J. 2003, \mnras, 339, 937

\bibitem[{{Greenhill} {et~al.}(2003{\natexlab{a}}){Greenhill}, {Booth},
  {Ellingsen}, {Herrnstein}, {Jauncey}, {McCulloch}, {Moran}, {Norris},
  {Reynolds}, \& {Tzioumis}}]{greenhill03}
{Greenhill}, L.~J., {Booth}, R.~S., {Ellingsen}, S.~P., {Herrnstein}, J.~R.,
  {Jauncey}, D.~L., {McCulloch}, P.~M., {Moran}, J.~M., {Norris}, R.~P.,
  {Reynolds}, J.~E., \& {Tzioumis}, A.~K. 2003{\natexlab{a}}, \apj, 590, 162

\bibitem[{{Greenhill} {et~al.}(2003{\natexlab{b}}){Greenhill}, {Booth},
  {Ellingsen}, {Herrnstein}, {Jauncey}, {McCulloch}, {Moran}, {Norris},
  {Reynolds}, \& {Tzioumis}}]{green03}
---. 2003{\natexlab{b}}, \apj, 590, 162

\bibitem[{{Greenhill} \& {Gwinn}(1997)}]{greenhill97}
{Greenhill}, L.~J. \& {Gwinn}, C.~R. 1997, \apss, 248, 261

\bibitem[{{Herrnstein} {et~al.}(1996){Herrnstein}, {Greenhill}, \&
  {Moran}}]{herrnstein}
{Herrnstein}, J.~R., {Greenhill}, L.~J., \& {Moran}, J.~M. 1996, \apjl, 468,
  L17+

\bibitem[{{Herrnstein} {et~al.}(1999){Herrnstein}, {Moran}, {Greenhill},
  {Diamond}, {Inoue}, {Nakai}, {Miyoshi}, {Henkel}, \& {Riess}}]{herrnstein99}
{Herrnstein}, J.~R., {Moran}, J.~M., {Greenhill}, L.~J., {Diamond}, P.~J.,
  {Inoue}, M., {Nakai}, N., {Miyoshi}, M., {Henkel}, C., \& {Riess}, A. 1999,
  \nat, 400, 539

\bibitem[{{Herrnstein} {et~al.}(2005){Herrnstein}, {Moran}, {Greenhill}, \&
  {Trotter}}]{herrnstein+05}
{Herrnstein}, J.~R., {Moran}, J.~M., {Greenhill}, L.~J., \& {Trotter}, A.~S.
  2005, \apj, 629, 719

\bibitem[{{Hobbs} \& {Nayakshin}(2009)}]{hobbs+nayakshin09}
{Hobbs}, A. \& {Nayakshin}, S. 2009, \mnras, 394, 191

\bibitem[{{Hofner} \& {Sparke}(1994)}]{hofner94}
{Hofner}, P. \& {Sparke}, L.~S. 1994, \apj, 428, 466

\bibitem[{{Hunter} \& {Toomre}(1969)}]{hunter69}
{Hunter}, C. \& {Toomre}, A. 1969, \apj, 155, 747

\bibitem[{{Jiang} \& {Binney}(1999)}]{jiang99}
{Jiang}, I.-G. \& {Binney}, J. 1999, \mnras, 303, L7

\bibitem[{{Kuijken}(1991)}]{Kuijken91}
{Kuijken}, K. 1991, \apj, 376, 467

\bibitem[{{Lai}(2003)}]{lai03}
{Lai}, D. 2003, \apjl, 591, L119

\bibitem[{{Lense} \& {Thirring}(1918)}]{lense1918}
{Lense}, J. \& {Thirring}, H. 1918, Physikalische Zeitschrift, 19, 156

\bibitem[{{Lu} {et~al.}(2009){Lu}, {Ghez}, {Hornstein}, {Morris}, {Becklin}, \&
  {Matthews}}]{lu+09}
{Lu}, J.~R., {Ghez}, A.~M., {Hornstein}, S.~D., {Morris}, M.~R., {Becklin},
  E.~E., \& {Matthews}, K. 2009, \apj, 690, 1463

\bibitem[{{Maloney} {et~al.}(1996){Maloney}, {Begelman}, \&
  {Pringle}}]{maloney96}
{Maloney}, P.~R., {Begelman}, M.~C., \& {Pringle}, J.~E. 1996, \apj, 472, 582

\bibitem[{{Martin}(2008)}]{martin08}
{Martin}, R.~G. 2008, \mnras, 387, 830

\bibitem[{{Miyoshi} {et~al.}(1995){Miyoshi}, {Moran}, {Herrnstein},
  {Greenhill}, {Nakai}, {Diamond}, \& {Inoue}}]{miyoshi95}
{Miyoshi}, M., {Moran}, J., {Herrnstein}, J., {Greenhill}, L., {Nakai}, N.,
  {Diamond}, P., \& {Inoue}, M. 1995, \nat, 373, 127

\bibitem[{{Natarajan} \& {Armitage}(1999)}]{natarajan99}
{Natarajan}, P. \& {Armitage}, P.~J. 1999, \mnras, 309, 961

\bibitem[{{Nayakshin}(2005)}]{nayakshin05}
{Nayakshin}, S. 2005, \mnras, 359, 545

\bibitem[{{Nelson} \& {Tremaine}(1995)}]{nelson95}
{Nelson}, R.~W. \& {Tremaine}, S. 1995, \mnras, 275, 897

\bibitem[{{Nenkova} {et~al.}(2002){Nenkova}, {Ivezi{\'c}}, \&
  {Elitzur}}]{nenkova02}
{Nenkova}, M., {Ivezi{\'c}}, {\v Z}., \& {Elitzur}, M. 2002, \apjl, 570, L9

\bibitem[{{Papaloizou} {et~al.}(1998){Papaloizou}, {Terquem}, \&
  {Lin}}]{papaloizou98}
{Papaloizou}, J.~C.~B., {Terquem}, C., \& {Lin}, D.~N.~C. 1998, \apj, 497, 212

\bibitem[{{Paumard} {et~al.}(2006){Paumard}, {Genzel}, {Martins}, {Nayakshin},
  {Beloborodov}, {Levin}, {Trippe}, {Eisenhauer}, {Ott}, {Gillessen}, {Abuter},
  {Cuadra}, {Alexander}, \& {Sternberg}}]{paumard06}
{Paumard}, T., {Genzel}, R., {Martins}, F., {Nayakshin}, S., {Beloborodov},
  A.~M., {Levin}, Y., {Trippe}, S., {Eisenhauer}, F., {Ott}, T., {Gillessen},
  S., {Abuter}, R., {Cuadra}, J., {Alexander}, T., \& {Sternberg}, A. 2006,
  \apj, 643, 1011

\bibitem[{{Petterson}(1977)}]{petterson77}
{Petterson}, J.~A. 1977, \apj, 216, 827

\bibitem[{{Phinney}(1989)}]{phinney89}
{Phinney}, E.~S. 1989, in NATO ASIC Proc. 290: Theory of Accretion Disks, ed.
  F.~{Meyer}, 457--+

\bibitem[{{Pringle}(1992)}]{pringle92}
{Pringle}, J.~E. 1992, \mnras, 258, 811

\bibitem[{{Pringle}(1996)}]{pringle96}
---. 1996, \mnras, 281, 357

\bibitem[{{Pringle}(1997)}]{pringle97}
---. 1997, \mnras, 292, 136

\bibitem[{{Sparke}(1984)}]{sparke84}
{Sparke}, L.~S. 1984, \apj, 280, 117

\bibitem[{{Sparke}(1986)}]{sparke86}
---. 1986, \mnras, 219, 657

\bibitem[{{Sparke} \& {Casertano}(1988)}]{sparke88}
{Sparke}, L.~S. \& {Casertano}, S. 1988, \mnras, 234, 873

\bibitem[{{Toomre}(1983)}]{toomre83}
{Toomre}, A. 1983, in IAU Symposium, Vol. 100, Internal Kinematics and Dynamics
  of Galaxies, ed. E.~{Athanassoula}, 177--185

\bibitem[{{Touma}(2002)}]{touma02}
{Touma}, J.~R. 2002, \mnras, 333, 583

\bibitem[{{Wu} {et~al.}(2008){Wu}, {Wang}, \& {Dong}}]{wu08}
{Wu}, S.-M., {Wang}, T.-G., \& {Dong}, X.-B. 2008, ArXiv e-prints, 807

\end{thebibliography}

\label{lastpage}

\end{document}